\documentclass[authorversion,sigconf]{acmart}

\usepackage{booktabs} 
\usepackage{multirow}
\usepackage{xcolor}
\usepackage{subcaption}
\usepackage{makecell}

\usepackage{soul}
\definecolor{myyellow}{rgb}{1,1,0}
\definecolor{mypink}{rgb}{1,0,1}
\usepackage{mathtools} \usepackage{xspace} 

\acmPrice{15.00}

\setcopyright{acmlicensed}
\acmJournal{PACMCGIT}
\acmYear{2025} \acmVolume{8} \acmNumber{1} \acmArticle{17} \acmMonth{5}\acmDOI{10.1145/3728295}

\setcitestyle{square}

\begin{document}


\title{LSNIF: Locally-Subdivided Neural Intersection Function}

\author{Shin Fujieda}
\affiliation{%
  \institution{Advanced Micro Devices, Inc.}
  \city{Tokyo}
  \country{Japan}}
\email{Shin.Fujieda@amd.com}
\orcid{0000-0002-2472-7365}

\author{Chih-Chen Kao}
\affiliation{%
  \institution{Advanced Micro Devices, Inc.}
  \city{Munich}
  \country{Germany}}
\email{ChihChen.Kao@amd.com}
\orcid{0000-0002-7631-2284}

\author{Takahiro Harada}
\affiliation{%
  \institution{Advanced Micro Devices, Inc.}
  \city{Santa Clara}
  \country{USA}}
\email{Takahiro.Harada@amd.com}
\orcid{0000-0001-5158-8455}

\renewcommand{\shortauthors}{Shin Fujieda, Chih-Chen Kao, and Takahiro Harada}

\begin{abstract}

Neural representations have shown the potential to accelerate ray casting in a conventional ray-tracing-based rendering pipeline. We introduce a novel approach called Locally-Subdivided Neural Intersection Function (LSNIF) that replaces bottom-level BVHs used as traditional geometric representations with a neural network.
Our method introduces a sparse hash grid encoding scheme incorporating geometry voxelization, a scene-agnostic training data collection, and a tailored loss function.  
It enables the network to output not only visibility but also hit-point information and material indices.
LSNIF can be trained offline for a single object, allowing us to use LSNIF as a replacement for its corresponding BVH.  
With these designs, the network can handle hit-point queries from any arbitrary viewpoint, supporting all types of rays in the rendering pipeline.
We demonstrate that LSNIF can render a variety of scenes, including real-world scenes designed for other path tracers, while achieving a memory footprint reduction of up to $106.2 \times$ compared to a compressed BVH. 

\end{abstract}

\ccsdesc[500]{Computing methodologies~Ray tracing}
\ccsdesc[500]{Computing methodologies~Neural networks}

\keywords{neural representations, multilayer perceptron, ray tracing}

\begin{teaserfigure}
    \setlength{\fboxrule}{0.002\linewidth}
    \setlength{\fboxsep}{0\linewidth}
    \setlength{\tabcolsep}{0.002\linewidth}
    \begin{tabular}{c@{\hspace{0.001\linewidth}}c@{\hspace{0.001\linewidth}}c}
    \textsc{Botanic Cornell Box} &
    \textsc{Statues} &
    \textsc{Classroom} \vspace{-0.175em} \\
    \includegraphics[width=0.32\textwidth]{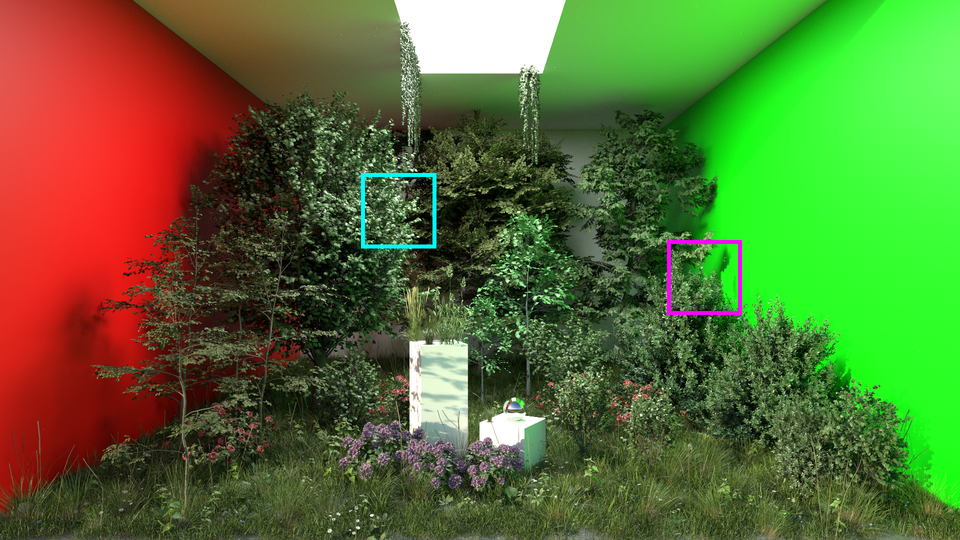} &
    \includegraphics[width=0.32\textwidth]{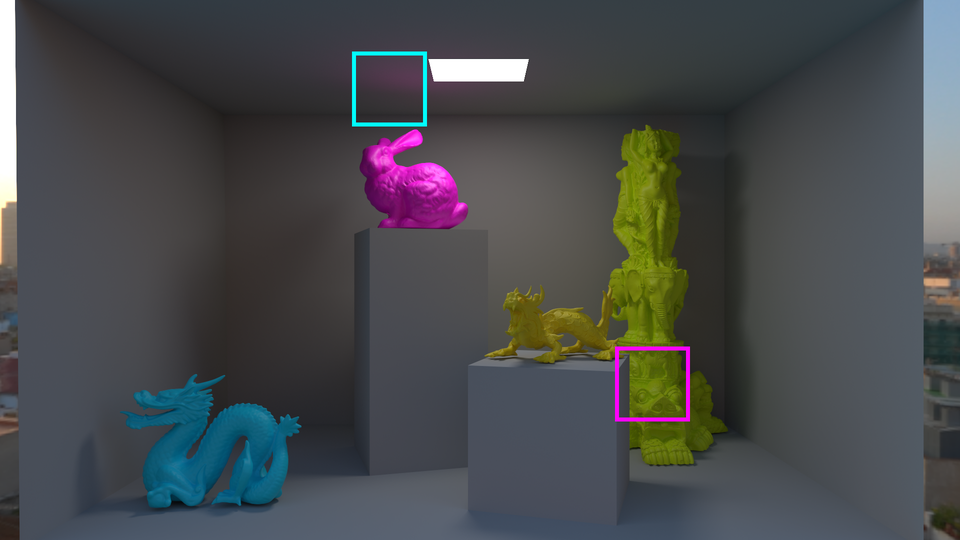} &
    \includegraphics[width=0.32\textwidth]{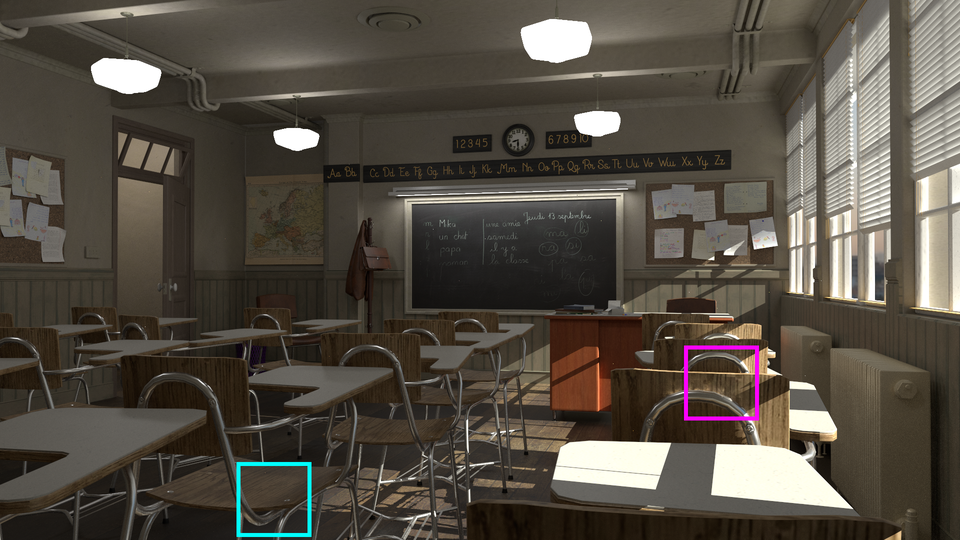} \vspace{-0.275em} \\
    \begin{tabular}{c@{\hspace{0.002\linewidth}}c}
        \includegraphics[width=0.159\textwidth]{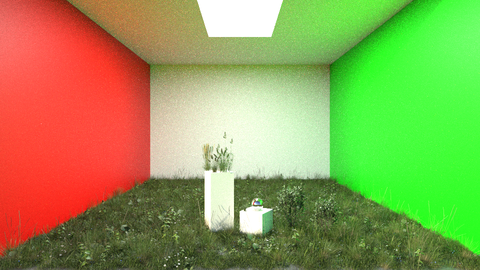} &
        \includegraphics[width=0.159\textwidth]{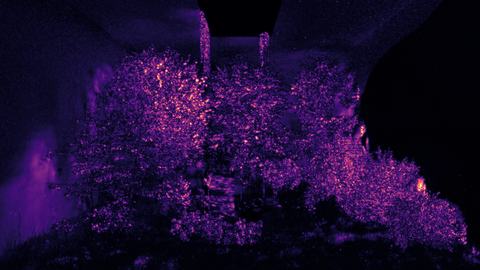}
    \end{tabular}
    &
    \begin{tabular}{c@{\hspace{0.002\linewidth}}c}
        \includegraphics[width=0.159\textwidth]{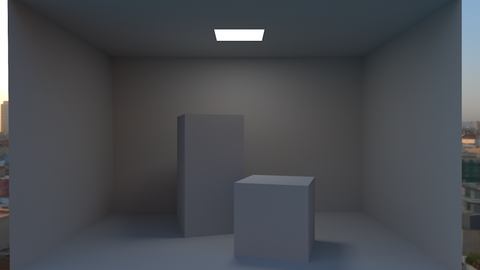} &
        \includegraphics[width=0.159\textwidth]{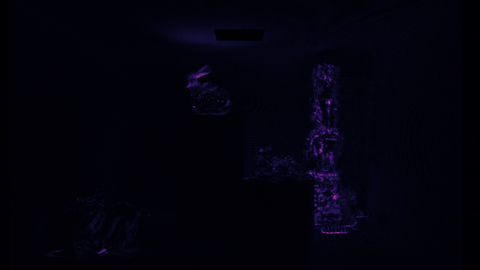}
    \end{tabular}
    &
    \begin{tabular}{c@{\hspace{0.002\linewidth}}c}
        \includegraphics[width=0.159\textwidth]{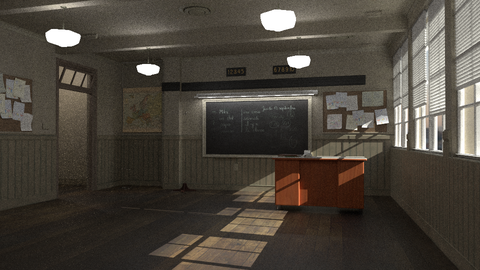} &
        \includegraphics[width=0.159\textwidth]{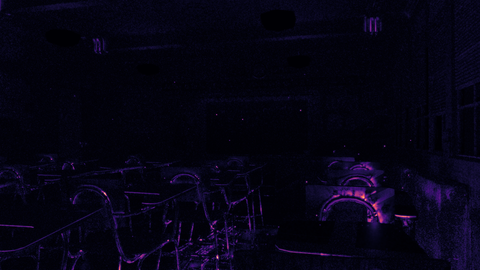}
    \end{tabular}
    \vspace{-0.275em}\\
    \hspace*{0.01em}
    \begin{tabular}{c@{\hspace{0.001\linewidth}}c@{\hspace{0.001\linewidth}}c}
        \fcolorbox{cyan}{cyan}{\includegraphics[width=0.102\textwidth]{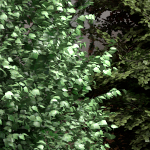}} &
        \fcolorbox{cyan}{cyan}{\includegraphics[width=0.102\textwidth]{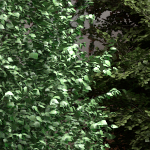}} &
        \fcolorbox{cyan}{cyan}{\includegraphics[width=0.102\textwidth]{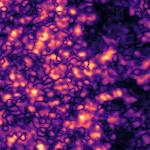}} \vspace{-0.275em}\\
        \fcolorbox{mypink}{mypink}{\includegraphics[width=0.102\textwidth]{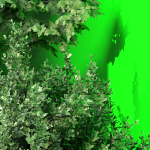}} &
        \fcolorbox{mypink}{mypink}{\includegraphics[width=0.102\textwidth]{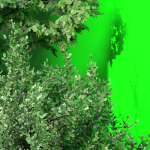}} &
        \fcolorbox{mypink}{mypink}{\includegraphics[width=0.102\textwidth]{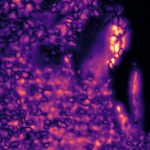}} \vspace{-0.275em}\\
        {\footnotesize LSNIF} &
        {\footnotesize Reference} &
        {\footnotesize FLIP: $0.108$} 
    \end{tabular}
    &
    \hspace*{0.01em}
    \begin{tabular}{c@{\hspace{0.001\linewidth}}c@{\hspace{0.001\linewidth}}c}
        \fcolorbox{cyan}{cyan}{\includegraphics[width=0.102\textwidth]{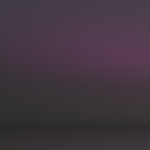}} &
        \fcolorbox{cyan}{cyan}{\includegraphics[width=0.102\textwidth]{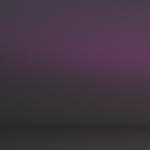}} &
        \fcolorbox{cyan}{cyan}{\includegraphics[width=0.102\textwidth]{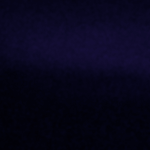}} \vspace{-0.275em}\\
        \fcolorbox{mypink}{mypink}{\includegraphics[width=0.102\textwidth]{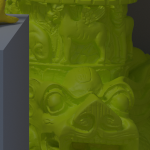}} &
        \fcolorbox{mypink}{mypink}{\includegraphics[width=0.102\textwidth]{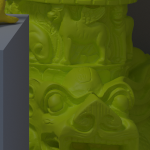}} &
        \fcolorbox{mypink}{mypink}{\includegraphics[width=0.102\textwidth]{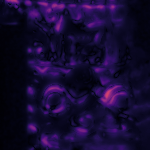}} \vspace{-0.275em}\\
        {\footnotesize LSNIF} &
        {\footnotesize Reference} &
        {\footnotesize FLIP: $0.021$}
    \end{tabular}
    &
    \hspace*{0.01em}
    \begin{tabular}{c@{\hspace{0.001\linewidth}}c@{\hspace{0.001\linewidth}}c}
        \fcolorbox{cyan}{cyan}{\includegraphics[width=0.102\textwidth]{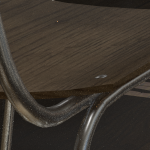}} &
        \fcolorbox{cyan}{cyan}{\includegraphics[width=0.102\textwidth]{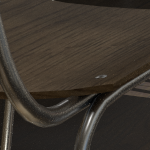}} &
        \fcolorbox{cyan}{cyan}{\includegraphics[width=0.102\textwidth]{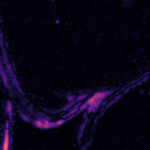}} \vspace{-0.275em}\\
        \fcolorbox{mypink}{mypink}{\includegraphics[width=0.102\textwidth]{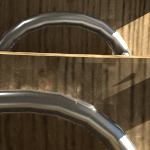}} &
        \fcolorbox{mypink}{mypink}{\includegraphics[width=0.102\textwidth]{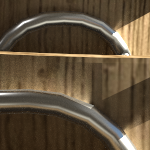}} &
        \fcolorbox{mypink}{mypink}{\includegraphics[width=0.102\textwidth]{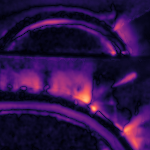}} \vspace{-0.275em}\\
        {\footnotesize LSNIF} &
        {\footnotesize Reference} &
        {\footnotesize FLIP: $0.036$}
    \end{tabular}
    \vspace{-0.275em}
    \end{tabular}
    \caption{Locally-Subdivided Neural Intersection Function (LSNIF) represents objects for which a bounding volume hierarchy (BVH) over triangles is usually used to reduce the algorithmic complexity of ray-geometry intersection. We show that LSNIF can be used to render complex scenes, foliage (\textsc{Botanic Cornell Box}), highly tessellated meshes containing 18.2 million polygons (\textsc{Statues}), and a scene designed for other path tracers with instancing and complex materials (\textsc{Classroom}). The images in the first row are rendered using LSNIF. The second row shows rendered images without LSNIF and the FLIP error between images rendered using LSNIF and the reference. The bottom two rows are close-up images. }
    \label{fig:teaser}
\end{teaserfigure}
\maketitle
\section{Introduction}

With the growing demand for high-quality rendering and fidelity, the importance of a physically based rendering pipeline continues to increase.
Physically based rendering relies on ray tracing to simulate light transport in 3D scenes~\cite{PBRT4e}.
Ray queries, fundamental operations in ray tracing, are typically accelerated using a bounding volume hierarchy (BVH) to search for ray-geometry intersections.
However, due to the irregular execution patterns during BVH traversal, such as branch execution and memory access divergence, it is difficult to achieve maximum efficiency on the SIMD architecture of GPUs.
Therefore, GPU vendors have added dedicated hardware to accelerate ray tracing~\cite{nvidiavolta, cdna}, but the workload is still expensive.

Although ray tracing using a BVH as the spatial acceleration structure is the de facto standard, researchers have been exploring possibilities using neural networks (NNs).
The two previously published approaches using NNs have major restrictions.
The Neural Intersection Function (NIF)~\cite{NIF} relies on overfitting neural networks to the current viewpoint, requiring online training that also depends on a BVH. From a memory usage perspective, while the neural networks are highly compressed, they still introduce a significant overhead to the memory footprint.
On the other hand, N-BVH~\cite{NBVH} eliminates the need for online training by pre-training the neural network for each scene. However, this approach cannot accommodate scene changes, which is its primary limitation.

In this paper, we present a novel approach that addresses the challenges of the previous works, extending its capability to cover broader functionality in the rendering pipeline.
We propose a new encoding scheme based on a sparse grid encoder and voxelization, which yields higher accuracy.
This encoding allows us to reduce the number of neural networks from two to one, decreasing the complexity of the algorithm compared to NIF.
Additionally, the proposed encoding scheme enables us to gather training data independent of the viewpoints and the lighting conditions in a scene, thus supporting arbitrary viewpoints or ray queries when adopted.
Finally, we design the loss and activation functions to extend the output of the neural network to report not only visibility but also auxiliary information, such as shading normals and material indices.
Consequently, the models can now support not only shadow-ray queries but also primary, secondary, or any other rays from deeper stages.
With these designs, we can pre-train the network models for individual objects and load the models to replace the corresponding parts of the BVH even in other applications.
We demonstrate that our method can handle practical situations by designing complex scenes in a 3D content creation tool, exporting these scenes to be trained using our approach, and loading the trained models into a separate application for rendering. We also show that LSNIF can be used as a building block for a path tracer that uses all four major functionalities in modern GPUs, such as rasterization, general computing, ray tracing hardware, and matrix cores. 

In summary, this paper makes the following contributions: 

\begin{itemize}
\item Novel Encoding Scheme: Introduces a new encoding scheme using a sparse grid encoder and voxelization, enhancing the accuracy of NN outputs. 

\item Scene-Independent Pre-trained Models: Enables training data collection independent of scene conditions, allowing support for arbitrary viewpoints, lighting conditions, and geometry transformations. The use of the model is not limited to occlusion rays. 

\item Extended Output Capabilities: Designs loss and activation functions to report visibility as well as additional information, such as normals and material indices, improving the versatility of the models and supporting various ray queries including shadow, primary, and deeper rays. 

\item Hybrid Rendering Pipeline: Combines three different techniques, such as rasterization, ray tracing, and machine learning, to construct a full path tracer. 

\end{itemize}
\section{Related Work}
\label{sec:related}

Ray tracing and path tracing are computationally intensive tasks that primarily involve finding intersections between rays and objects. To expedite this process, a spatial acceleration data structure like a BVH is typically used, reducing the time complexity from linear to logarithmic. Given the embarrassingly parallel nature of ray tracing, GPUs can be leveraged by assigning individual rays to each GPU thread. However, branch execution and memory access divergence can reduce performance, making it difficult to achieve maximum efficiency~\cite{kao2018exploring}. Although specialized hardware has been developed to accelerate execution, it still suffers from memory access divergence or leads to higher memory consumption, presenting ongoing challenges~\cite{meister2021survey}. There have been studies that approximate geometries with simpler representations such as geometric levels of details (LODs)~\cite{Christensen2003, Yoon2006, razor}, voxels~\cite{ikeda2022multi, zeng2023ray}, and distance fields~\cite{10.1145/3550340.3564231}.

With recent advances in NN or machine learning, using NN in ray tracing or rendering has been explored. These are classified into two categories: one is to use NN as a post-process to upscale~\cite{shi2016real}, denoise~\cite{zhang2024neural}, or generate frames~\cite{briedis2021neural}. The other category is to embed NN in the rendering pipeline itself to replace textures~\cite{10.1145/3450626.3459795, 10.1145/3592407,Fujieda2024NeuralTB} or materials~\cite{10.1145/3659577}, or to use it as a cache~\cite{10.1145/3450626.3459812} or for importance sampling~\cite{10.1145/3341156}. As multilayer perceptrons (MLPs) are not well-suited for reconstructing high-frequency signals, various input encoders are often used in conjunction, such as positional encoding~\cite{10.1145/3503250}, grid encoding~\cite{10.1145/3450626.3459795, mueller2022instant}, and hybrids of these methods~\cite{10.1145/3592407, 10.2312:pg.20231273}. Some approaches incorporate geometric information~\cite{takikawa2021nglod}, similar to the method we use in our paper. 
Some researchers have also explored replacing BVHs with NNs. For example, Fujieda et al. introduced the Neural Intersection Function (NIF), the first work to use two NNs to replace the bottom-level BVH traversal to query visibility status~\cite{NIF}. By utilizing grid encoding and MLPs, NIF replaces the most divergent part of the computation with matrix multiplications, which are regular algorithms that can be accelerated using hardware components such as AMD Matrix Cores, Tensor Cores, and Wave Matrix Multiply Accumulate (WMMA) instructions~\cite{schieffer2024rise,cui2024acceleration}.
NIF uses online training for the current viewpoint of the scene. Although it allows for applying it for dynamic scenes with re-training, it cannot completely remove BVHs for the geometries as they are required to generate training data. Also, they limit the use of NIF only for shadow rays. Weier et al. proposed N-BVH with a neural ray-query pipeline that samples uniformly along a given ray-box intersection interval to collect feature vectors similar to the sampling employed in NeRF~\cite{10.1145/3503250}. These feature vectors are concatenated and fed to the MLP to obtain reconstructed signals. Although N-BVH does not require online training, the models can only be trained on a per-scene basis.
This restriction prevents further modifications such as object transformation and any interaction such as dynamically adding objects.

Unlike previous methods, the network models in LSNIF are designed to be trained for each object. LSNIF employs a uniform sampling strategy to cast rays for collecting training data, independent of viewpoint position or scene lighting. As a result, each object can be trained individually, and the grid encoding, MLP, and voxels are stored in binary format after offline training. This allows the LSNIF models to be trained offline, and then loaded into another application. Moreover, the models in LSNIF can output properties including visibility, hit points, normals, and material information, supporting not only shadow rays, but also primary, secondary, and all rays from deeper stages of path tracing, thus supporting a wide range of applications.

\begin{figure*}
     \centering
     \includegraphics[width=\textwidth]{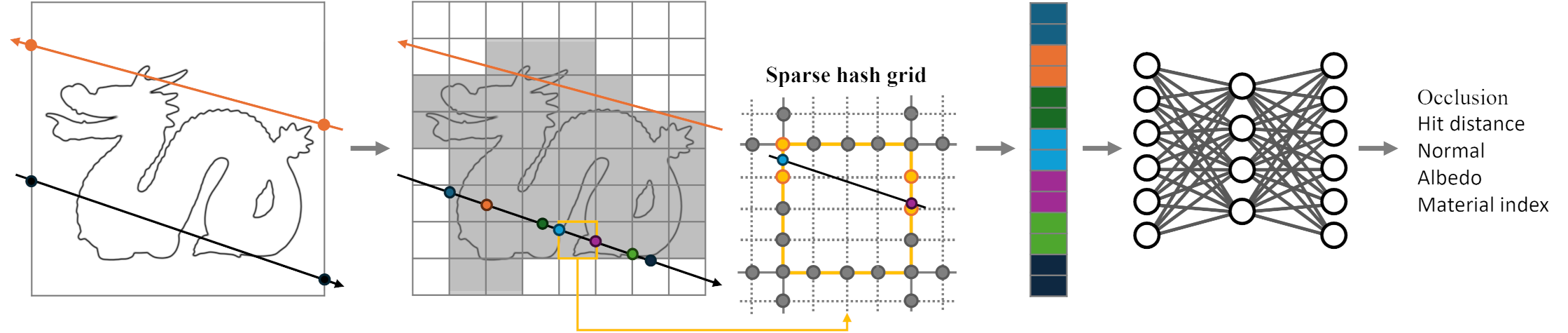}
     \caption{The illustration of LSNIF methodology. First, the intersection points of a ray with the object's AABB are computed. These points are then used to perform DDA against the voxels, followed by the calculation of hit points on the surfaces of these voxels. The hit points are processed using 3D sparse hash grid encoding, with interpolated feature vectors concatenated into a large vector. This vector is then fed into the MLP which outputs the intersection information of the ray with the geometry.}
     \label{fig:3:flow}
\end{figure*}

\section{Locally-Subdivided Neural Intersection Function} \label{sec:lsnif}
Our method addresses the limitations of existing works discussed in Sec.~\ref{sec:related} by extending NIF.
We propose a novel neural representation optimized for a single object, called Locally-Subdivided Neural Intersection Function (LSNIF).
Fig.~\ref{fig:3:flow} illustrates the overview of the LSNIF methodology. 

\subsection{Local Geometry Voxelization}

The uniqueness of the input to the neural network is crucial for it to effectively learn the underlying function, such as mapping $3$D input positions to the corresponding shape information.
Our objective is to infer the geometric properties of an object at the intersection point between a ray and the geometry using NNs. To achieve this, each input feature must uniquely represent a specific ray.
Rays with the same direction but originating from different points along that direction are distinct yet converge at a single intersection point. This can lead to confusion for the network, a phenomenon known as ray aliasing.

NIF addresses the ray-aliasing by using the ray direction and the intersection point between the ray and the AABB or the surface of the geometry as the input instead of the ray origin.
While this approach is effective for online training, where the network is trained with a limited set of rays generated from a single viewpoint under fixed lighting conditions, it is not suitable for dynamic content. In scenarios with moving viewpoints or changing lighting conditions, the network must be pre-trained with arbitrary rays cast from any position and in any direction, presenting significant challenges.
Furthermore, concatenating input features from multiple $2$D grid encodings, as done in NIF for position and direction, hinders small MLPs from accurately inferring specific geometric properties.
This is because different rays, such as those with the same origin but different directions, share identical features extracted from the ray origin, leading to the aliasing problem. 
One possible solution to make the input features unique is to use $4$D grid encoding to represent distinct features for each pair of position and direction. However, this approach is computationally expensive and memory-intensive, making it impractical for real-world applications.

Another option to improve prediction accuracy is to provide more data as input to the NN.
In this work, we adopt this strategy by voxelizing the geometry into a low-resolution grid in local space. The intersection points between a ray and the voxels are then used as input to the network. (Fig.~\ref{fig:3:flow}). This approach enhances the variation in the inputs, enabling the network to more effectively distinguish between inputs with different features and reduce the impact of aliasing. As a result, the network can more accurately reconstruct the original geometry.

Since our focus is on the location of the geometry's surface, only the surface polygons are voxelized. Furthermore, because our method makes no assumptions about the topology of the geometry, it is fully compatible with any polygon soup.  The additional size of this voxel data is negligible compared to other stored model data, as each voxel only contains binary occupancy information.

For a ray intersecting the AABB of geometry, rather than using the hit point and direction as inputs to the neural network, we examine the intersections with the voxels as follows:
we first extract the voxels hit by the ray using the digital differential analyzer (DDA) algorithm ~\cite{VoxelTraversal}.
Then, we compute the intersection points to the voxel boundaries in the local space of the geometry with a simple ray-box intersection test as shown in the middle of Fig.~\ref{fig:3:flow}.
The resulting list of intersection points is then used as input to the neural network.
Our approach considers the underlying geometry by sampling positions closer to the surface, unlike methods that sample at regular intervals.~\cite{NBVH}.

This enables the network to learn features from more correlated input points, as demonstrated later. Furthermore, in contrast to regular interval sampling, our approach ensures that each point intersects an occupied voxel, eliminating the need for explicit voxel occupancy checks and improving the distinguishability of features from each ray direction.

\subsection{Sparse Hash Grid Encoding} \label{sec:sparse_grid}
Our input to the NN is the intersection points for a given ray and the voxels of the local geometry.
We encode each intersection point into a feature vector with a multi-resolution hash grid to represent sparse and complex $3$D signals compactly~\cite{mueller2022instant}.
Then, we concatenate the feature vectors into a single large vector to reduce the number of inference queries and feed it into the NN.
Because concatenating the feature vectors on voxel boundaries implicitly distinguishes whether a ray origin lies inside or outside the AABB, we do not need to handle the inside and outside cases separately, as NIF requires.
Most feature vectors on the grid vertices are unnecessary and simply a waste of memory as all our input points are located on the voxel boundaries.
Therefore, we propose to use a sparse hash grid tailored for the distribution of the input points.

Sparse hash grid encoding is a specialized version of the multi-resolution hash grid that stores the feature vectors only on the voxel boundaries, as shown in the middle of Fig.~\ref{fig:3:flow}, where this encoding structure is described in $2$D for simplicity.
We ignore grid vertices that do not lie on the voxel boundaries and store only the feature vectors on the remaining vertices, which reduces the memory consumption of the grid encoder.
Additionally, this encoding structure decreases the number of queries to the grid vertices, which is beneficial for the inference speed.
With the $2$D multi-resolution hash grid, we need to query at four grid vertices bilinearly interpolated to get the feature vector for a given position.
However, with the $2$D sparse hash grid, we only need to query at most two grid vertices linearly interpolated because the feature vectors are stored only on the voxel boundaries, which results in a 50$\%$ reduction of memory access.
In the $3$D case, we need to query eight grid vertices with the dense hash grid, but only four with the sparse hash grid.
Therefore, this sparse structure is more efficient in terms of memory footprint and inference speed than the dense one.
Since the sparse structure of the grid makes it challenging to explicitly access feature vectors associated with the corresponding hit points, we therefore rely on a hash table to map the index of the grid vertex to the memory address where the feature vector is stored.

\begin{figure*}
     \centering
     \begin{subfigure}[b]{0.16\textwidth}
         \centering
         \includegraphics[width=\textwidth]{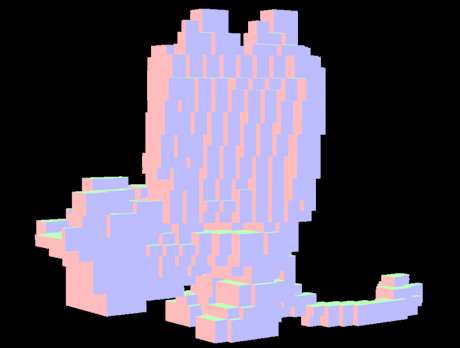}
         \caption{Voxels}
     \end{subfigure}
     \hfill
     \begin{subfigure}[b]{0.16\textwidth}
         \centering
         \includegraphics[width=\textwidth]{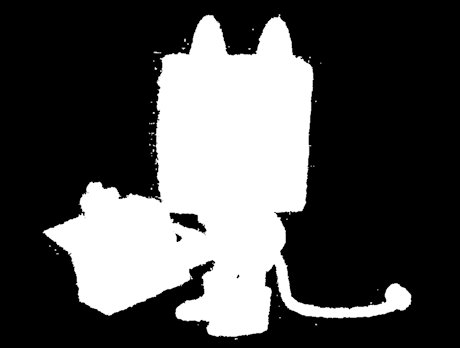}
         \caption{Occlusion}
     \end{subfigure}
     \hfill
     \begin{subfigure}[b]{0.16\textwidth}
         \centering
         \includegraphics[width=\textwidth]{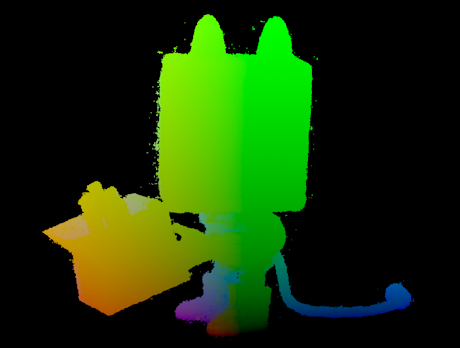}
         \caption{Hit distance}
     \end{subfigure}
     \hfill
     \begin{subfigure}[b]{0.16\textwidth}
         \centering
         \includegraphics[width=\textwidth]{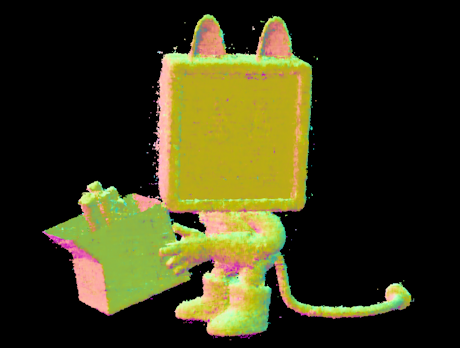}
         \caption{Normal}
     \end{subfigure}
     \hfill
     \begin{subfigure}[b]{0.16\textwidth}
         \centering
         \includegraphics[width=\textwidth]{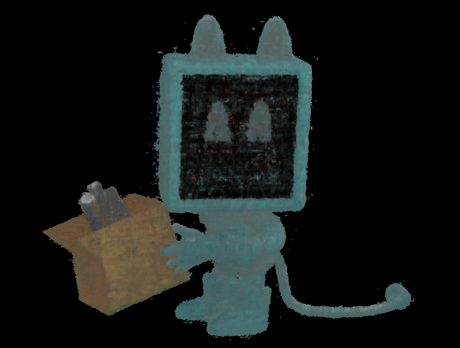}
         \caption{Albedo}
     \end{subfigure}
     \hfill
     \begin{subfigure}[b]{0.16\textwidth}
         \centering
         \includegraphics[width=\textwidth]{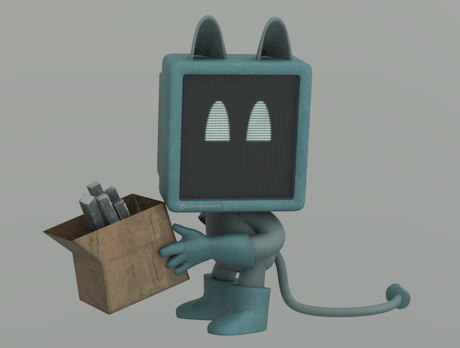}
         \caption{Rendered}
     \end{subfigure}
        \caption{(a) Voxel representation of the geometry. (b, c, d, e) Outputs from LSNIF for primary rays. (f) Path-traced image with LSNIF. Note that (f) utilizes rasterization for primary visibility instead of the LSNIF representation shown in (b, c, d, e).}
        \label{fig:a:nifoutputs}
\end{figure*}

\subsection{Network Design}
We use an MLP to infer the geometric properties of each object for a ray.
A single MLP is trained for each object with the $3$D sparse hash grid encoding described in Sec.~\ref{sec:sparse_grid} and takes the concatenated feature vectors from the sparse hash grid as input.
The MLP then predicts multiple geometric properties simultaneously, including occlusion, hit distance, normal, albedo, and material index, as shown in Fig.~\ref{fig:a:nifoutputs}.
This approach is more efficient than using separate MLPs for each property in terms of memory usage and inference speed.
In addition, the prediction of a material index makes it possible to assign multiple materials to a single object, which is required to represent complex and realistic scenes.

The components on LSNIF include the voxel data, the sparse hash grid, and the MLP for each object.
All of them can be pre-computed and pre-trained. The generated voxel data, optimized grid buffer, and neural network weights are stored on disk. At runtime, they are loaded as LSNIF objects, eliminating the need for bottom-level BVH traversals.

\section{Path Tracing Pipeline with LSNIF}
We adopt and integrate LSNIF in a wavefront path tracer. The challenge of integrating LSNIF lies in the fact that intersecting with an LSNIF object requires inference execution of the NN. Since the scene typically contains many LSNIF objects, each with potentially different configurations and NN parameters, this introduces code and data divergence. Minimizing this divergence is crucial for achieving optimal performance, especially on GPUs.

\begin{figure*}
\centering
  \setlength{\fboxrule}{0.002\linewidth}
  \setlength{\fboxsep}{0\linewidth}
  \setlength{\tabcolsep}{0.002\linewidth}
    \begin{tabular}{c@{\hspace{0.002\linewidth}}c@{\hspace{0.002\linewidth}}c@{\hspace{0.002\linewidth}}c@{\hspace{0.002\linewidth}}}
        \raisebox{0.85\height}[0pt][0pt]{
        \begin{tabular}{c@{\hspace{0.002\linewidth}}c@{\hspace{0.002\linewidth}}}
        \includegraphics[width=0.113\textwidth]{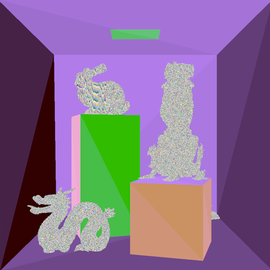} &
        \includegraphics[width=0.113\textwidth]{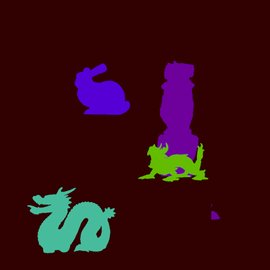} \vspace{-0.175em}\\
        \includegraphics[width=0.113\textwidth]{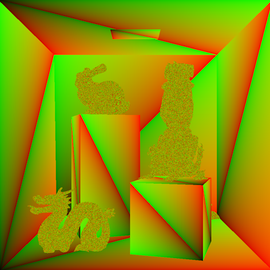} &
        \includegraphics[width=0.113\textwidth]{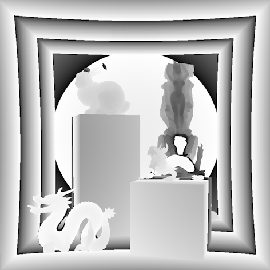} \vspace{-0.175em}\\
        \end{tabular}
        }
     &
    \includegraphics[width=0.24\textwidth]{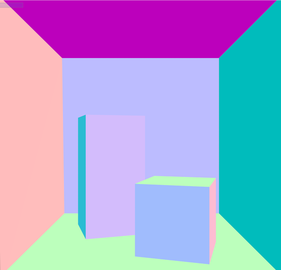} &
    \includegraphics[width=0.24\textwidth]{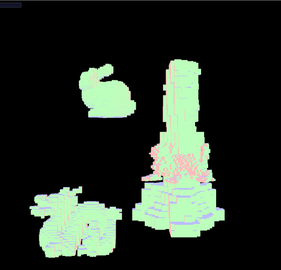} &
    \includegraphics[width=0.24\textwidth]{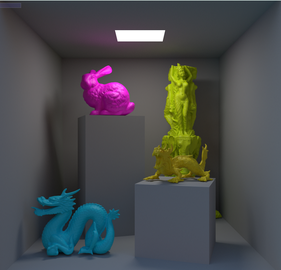} \vspace{-0.175em}\\
    {\footnotesize (a)} &
    {\footnotesize (b)} &
    {\footnotesize (c)} &
    {\footnotesize (d)}
    \end{tabular}
        \caption{Visual illustration of the proposed rendering pipeline. (a) the scene is rasterized from the camera. G-buffers are used to generate rays from the primary vertices. (b) the rays are intersected with triangles of non-LSNIF objects using BVHs. Then (c) the rays are tested against LSNIF objects which store a coarse voxel representation of the geometry. (d) the remaining part of the rendering pipeline stays the same to render global illumination.}
        \label{fig:pipeline}
\end{figure*}

The following outlines the techniques and strategies we adopted to integrate LSNIF into a wavefront path tracer, aiming to optimize performance:
Path tracing starts by shooting primary rays from the camera.
Although LSNIF is designed to support all types of rays, its application to primary rays results in lower accuracy compared to traditional BVH traversal, as shown in Fig.~\ref{fig:a:nifoutputs} (b) - (e).
Therefore, we restrict the application of LSNIF to non-primary rays in this paper.
However, resolving the intersection of primary rays using ray tracing requires a BVH, which defeats the purpose of LSNIF in terms of memory savings.
We propose to address this problem by leveraging rasterization to create G-Buffers storing shape indices, primitive indices, UV coordinates, and camera distances, replacing primary ray hit information as shown in Fig.~\ref{fig:pipeline} (a).
This approach requires retaining vertex and face data for rasterization but eliminates the need for BVH construction and storage for objects replaced with LSNIF.

LSNIF can be used as a substitute for the bottom-level BVH traversal for rays other than the primary rays used in path tracing.
Other components in the path tracer do not need to be modified and remain the same.
Similar to NIF, applying LSNIF to objects with a small number of triangles is inefficient. Because of that, we can set a condition measuring the complexity, e.g., setting a threshold to the number of faces, to switch between LSNIF and BVH traversal.
For non-LSNIF objects, we use a two-level BVH where the top-level BVH organizes a set of bottom-level BVHs, each of which contains the geometry of a single object.
For non-primary rays, we first check the intersections with triangles for non-LSNIF objects using a two-level BVH and then find the intersection points between rays and the voxels of LSNIF objects.
To find the intersection points against LSNIF objects efficiently, we construct a dedicated top-level BVH for LSNIF objects (LSNIF BVH) and use a two-phase approach: a broad phase followed by a narrow phase.
In the broad phase, we traverse the LSNIF BVH to achieve the potentially intersecting list of sets of ray information and the index of the LSNIF object. The simplest implementation would be executing the intersection tests to LSNIF in the BVH traversal, but it introduces data and code divergence which have a negative impact on the performance.
To improve efficiency, we store the ray-LSNIF object index pairs in the list, which is sorted by the index of the LSNIF object to maximize the coherence of the inference execution.
In the narrow phase, we use the DDA-based algorithm to extract the voxels hit by the ray stored in the potentially intersecting list and then compute hit points between the ray and the voxels.
Finally, we use the hit points as input to the LSNIF model, allowing the model to infer the geometric properties instead of traversing the bottom-level BVHs. If a hit is detected on the LSNIF object, the ray's hit information is updated with the model's outputs, which are then used in the subsequent steps of the rendering pipeline, as shown in Fig.~\ref{fig:pipeline}. Our current implementation splits the ray-triangle and ray-LSNIF intersection tests into two phases and executes the inference for all the collected ray-LSNIF object index pairs in a single GPU dispatch. While this approach may not be the most optimal in terms of computational complexity, it is well-aligned with the constraints and capabilities of modern GPU architectures and programming models. Future research will aim to enhance performance while minimizing the impact on GPU execution, considering hardware limitations such as branch or memory access divergence and the inability to independently query results from the neural network during ray processing.

\section{Implementation Details}
We integrated LSNIF into a HIP-based~\cite{hip} wavefront path tracer with the support of the hardware ray tracing cores on RDNA3 GPU~\cite{navi3}.
Rasterization for primary rays was implemented with OpenGL, and we utilize HIP-OpenGL interoperability to enable our HIP kernels to directly read the necessary hit information of primary rays from the OpenGL-generated textures.
For training and inference of the NN on GPUs, we implemented using C++ and HIP.
Our implementation is to fuse operations from all layers into a single kernel to minimize the overhead of memory access and kernel launch.
All the computations in the training and inference procedures, except for the gradient accumulation, are executed with half-precision floating points.
We used the Wave Matrix Multiply Accumulate (WMMA) instructions~\cite{schieffer2024rise} to accelerate them. 

\subsection{Architecture} \label{sec:arch}
We use a small MLP with two hidden layers, each with $128$ neurons and leaky ReLU activation functions, for all the experiments we show in this paper.
The output layer has different activation functions depending on the output signals.
We use the sigmoid activation function for occlusion, hit distance, and albedo, the linear activation function (identity function) for normal, and the softmax activation function for material index.
The dimension of the input vector depends on the number of intersection points for a given ray and the voxels.
However, to fix the size of the input layer, we impose an upper bound, $H$, on the number of intersection points.
In this work, we empirically set $H = 18$ and fill the remaining vector elements with zeros.
We discuss this upper bound, $H$, in more detail in Section.~\ref{sec:parameter}.
Also, the number of neurons in the output layer is determined by the number of materials attached to the object, i.e. $8 + N_{mat}$.
Note that we need to re-train the NN with the new output layer size for the object if $N_{mat}$ is changed.
The NN weights are initialized with the Kaiming initialization~\cite{He_2015_ICCV}.

Sparse hash grid encoding comprises two levels of grids with resolutions from $64^3$ to $128^3$.
The voxel resolution for each geometry is $32^3$.
The feature vectors on the sparse grid are stored only on the voxel boundaries, allowing us to use a relatively small hash-map size of $2^{17}$.
Each entry in the hash map stores a $3$D feature vector initialized with the uniform distribution $\mathcal{U}(-10^{-4}, 10^{-4})$.
A detailed analysis of hash-map sizes and voxel resolutions is provided in Sec.~\ref{sec:parameter}.
With these parameters, the memory footprint of a single LSNIF is only about $1.56$ MB, where $4$ KB, $1,536$ KB, and about $62$ KB are used for voxels, the input encoder, and the NN.
This memory footprint remains constant regardless of the complexity of the object geometry.
So, the efficiency of LSNIF increases as the complexity of the object grows.

\subsection{Training}
We train LSNIF for a single object with the training data prepared using balanced sampling with $50\%$ of the rays originating from outside the AABB and $50\%$ from the object surface.
This ensures that LSNIF can learn the representation of both ray types equally.
To generate rays originating outside the AABB, we first uniformly sample points on a sphere that encloses the AABB. These points serve as ray origins.
Ray directions are then determined via cosine-weighted hemispherical sampling oriented toward the center of the sphere.
On the other hand, rays originating on the object's surface are generated by first casting rays from outside the AABB, sampled with the previously described process, to find intersection points on the surface.
From these hit points, ray directions are sampled using cosine-weighted hemispherical sampling aligned with the surface normals.
The outputs are compared against the ground truth values on the target object, obtained via ray casting using a BVH.
With the training data prepared, we jointly optimize the NN and the sparse hash grid encoding in batches of $2^{18}$ rays using the Adam optimizer with a learning rate of $0.01$~\cite{Adam}.

\subsection{Activation and Loss Functions}
The MLP in LSNIF infers multiple signals at once, including occlusion, hit distance, shading normal, albedo, and material index.
We describe the activation and loss functions used for each signal below.

\paragraph{Occlusion}
The occlusion is a binary value that indicates whether the ray is occluded or not.
We can consider this as a binary classification problem, so we use the sigmoid activation function and the binary cross-entropy loss for this signal.
The output probability is then thresholded at $0.5$ to determine the final occlusion value.
To avoid learning irrelevant features, we set the losses for non-occlusion signals to zero when the reference occlusion is zero.

\paragraph{Local Hit Distance}
Although the hit distance represents the distance from the ray origin to the intersection point, we learn the local distance along the ray-AABB intersection interval, which simplifies the signal representation to a $1$D continuous value in the range $[0, 1]$.
We use the sigmoid activation function and the mean absolute error (MAE) loss, which is more robust to outliers than the mean squared error (MSE) loss.

\paragraph{Shading Normal}
The normal is a $3$D vector representing the surface normal at the intersection point.
We use the linear activation function (identity function) and the cosine similarity loss for this signal, which empirically shows better accuracy than other loss functions, such as the MAE and the MSE losses.

\paragraph{Albedo}
The albedo is a $3$D vector representing the surface's base color at the intersection point.
We use the sigmoid activation function and the relative L2 loss~\cite{noise2noise}, which shows better results than the MSE loss for this signal.
Other BSDF parameters are determined by looking up the attached material with the predicted material index.
We did not infer texture coordinates, which we found difficult to predict because they are too sensitive to a small error, so LSNIF only supports albedo textures on materials.

\paragraph{Material Index}
The material index is an integer value indexing one material attached to the object.
This enables the object to have multiple materials with different properties, which is necessary for realistic rendering.
We can consider this as a multi-class classification problem, so we use the softmax activation function and the categorical cross-entropy loss.

\section{Experimental Results}

\begin{figure*}[tb]
  \setlength{\fboxrule}{0.002\linewidth}
  \setlength{\fboxsep}{0\linewidth}
  \setlength{\tabcolsep}{0.002\linewidth}
  \begin{tabular}{c@{\hspace{0.002\linewidth}}c@{\hspace{0.003\linewidth}}c@{\hspace{0.002\linewidth}}c@{\hspace{0.003\linewidth}}c@{\hspace{0.002\linewidth}}c@{\hspace{0.003\linewidth}}c@{\hspace{0.002\linewidth}}c@{\hspace{0.003\linewidth}}c@{\hspace{0.002\linewidth}}c@{\hspace{0.003\linewidth}}c@{\hspace{0.002\linewidth}}c}

  \multicolumn{2}{c}{{\footnotesize $V = 2^3$}} &
  \multicolumn{2}{c}{{\footnotesize $V = 4^3$}} &
  \multicolumn{2}{c}{{\footnotesize $V = 8^3$}} &
  \multicolumn{2}{c}{{\footnotesize $V = 16^3$}} &
  \multicolumn{2}{c}{{\footnotesize $V = 32^3$}} &
  \multicolumn{2}{c}{{\footnotesize $V = 64^3$}} \vspace{-0.205em}\\

  \includegraphics[width=0.08\textwidth]{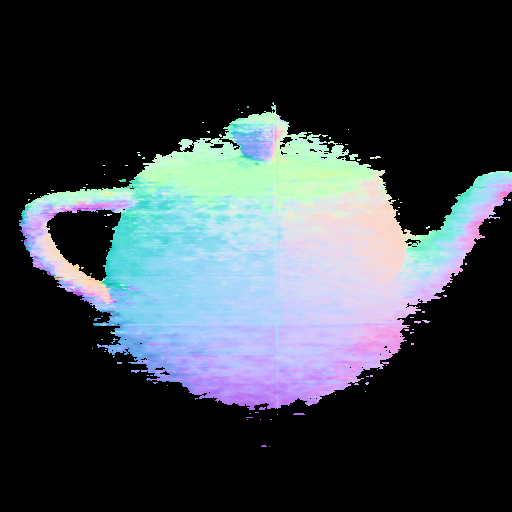} &
  \includegraphics[width=0.08\textwidth]{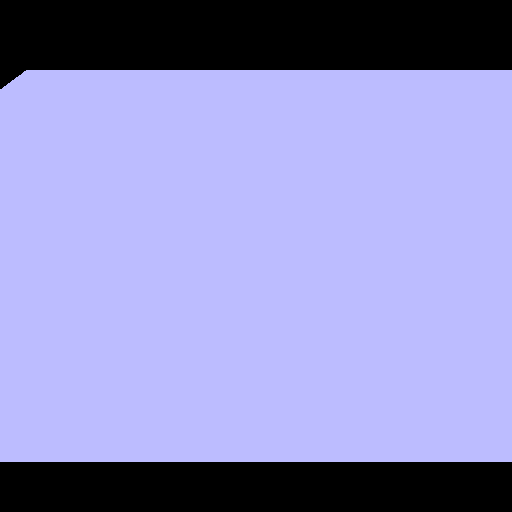} &

  \includegraphics[width=0.08\textwidth]{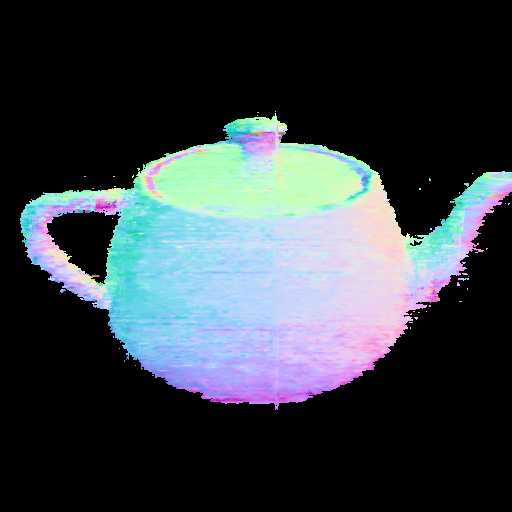} &
  \includegraphics[width=0.08\textwidth]{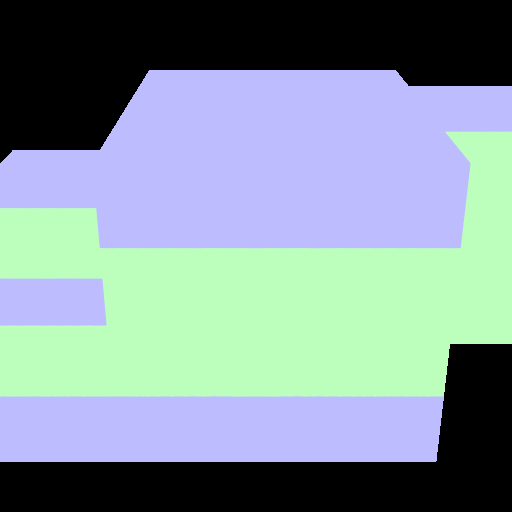} &

  \includegraphics[width=0.08\textwidth]{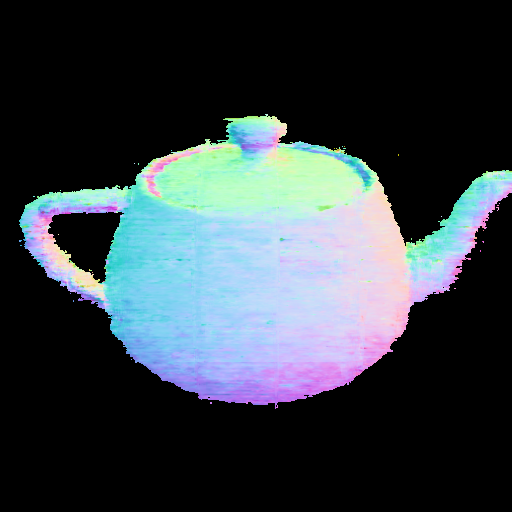} &
  \includegraphics[width=0.08\textwidth]{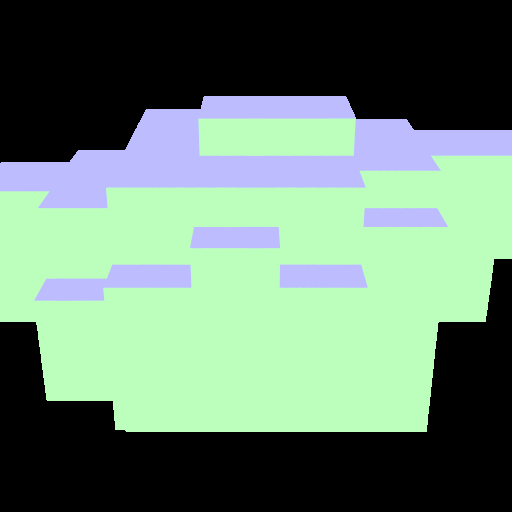} &

  \includegraphics[width=0.08\textwidth]{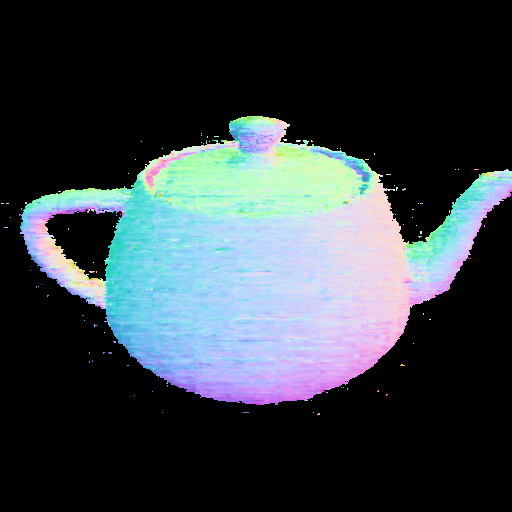} &
  \includegraphics[width=0.08\textwidth]{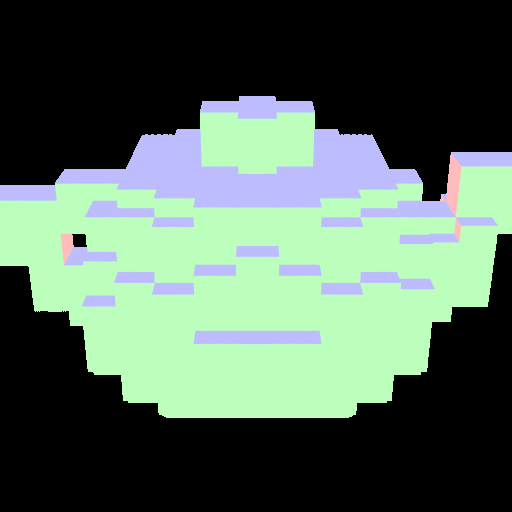} &

  \includegraphics[width=0.08\textwidth]{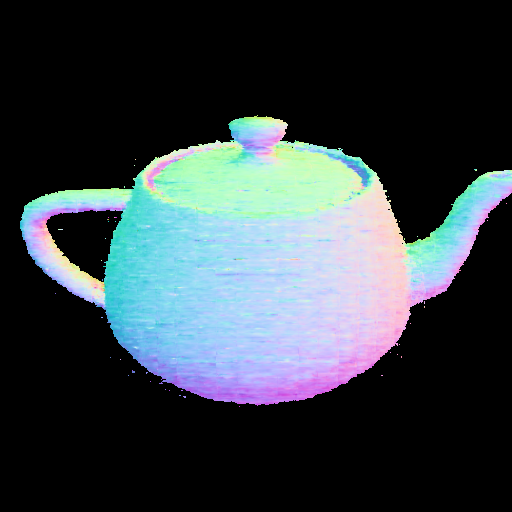} &
  \includegraphics[width=0.08\textwidth]{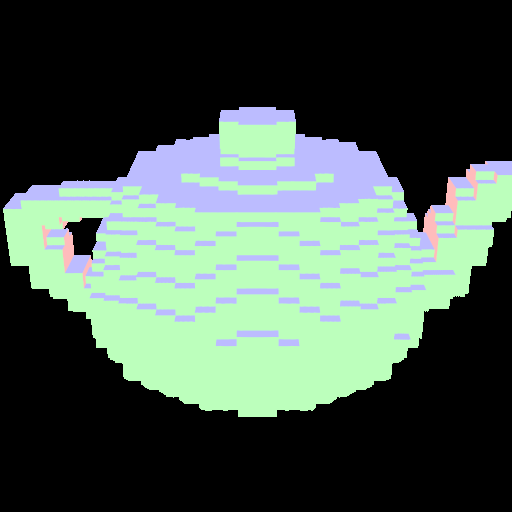} &

  \includegraphics[width=0.08\textwidth]{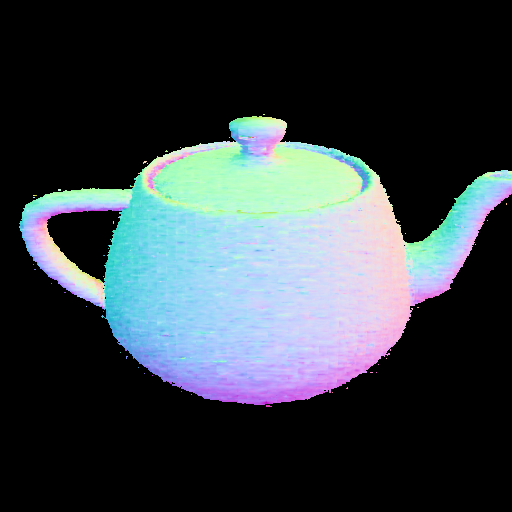} &
  \includegraphics[width=0.08\textwidth]{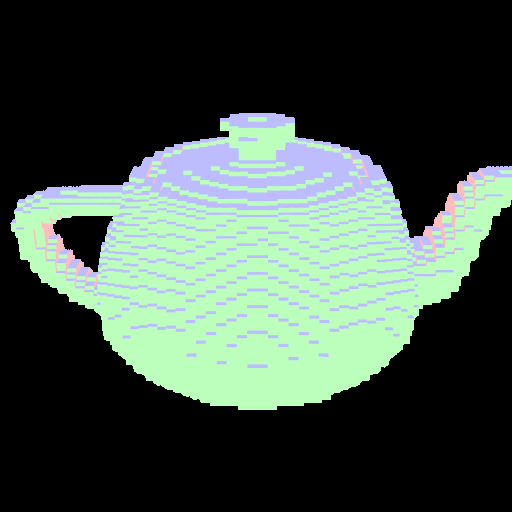} \vspace{-0.275em}\\

  \includegraphics[width=0.08\textwidth]{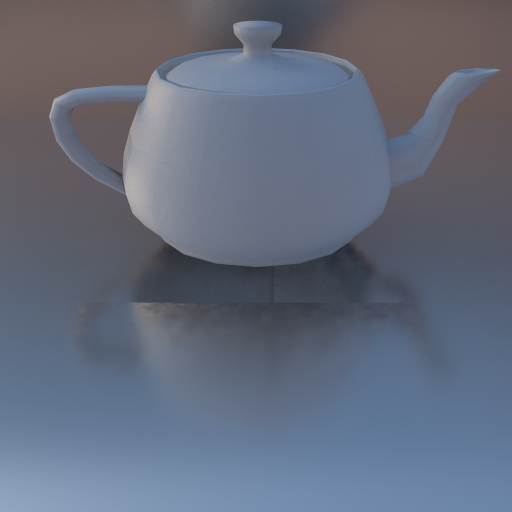} &
  \includegraphics[width=0.08\textwidth]{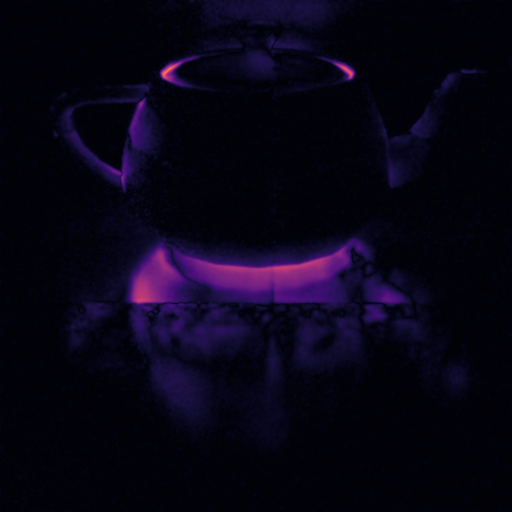} &

  \includegraphics[width=0.08\textwidth]{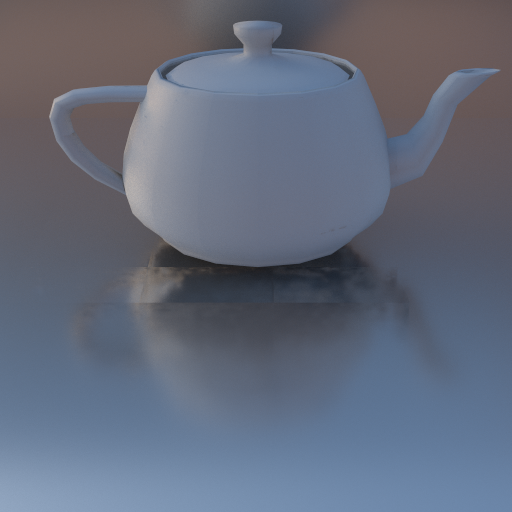} &
  \includegraphics[width=0.08\textwidth]{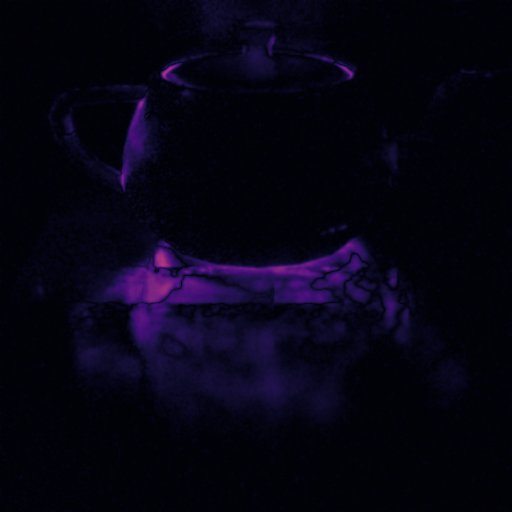} &

  \includegraphics[width=0.08\textwidth]{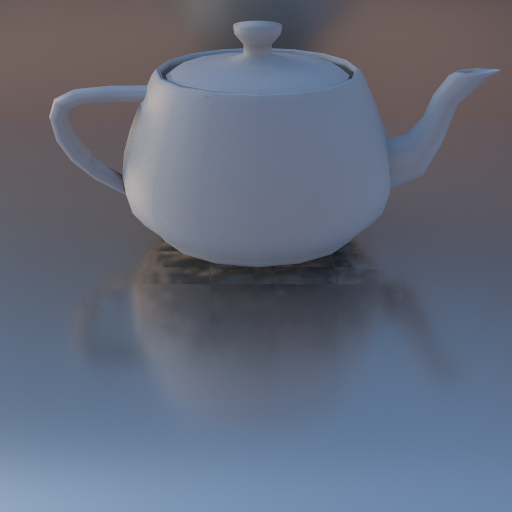} &
  \includegraphics[width=0.08\textwidth]{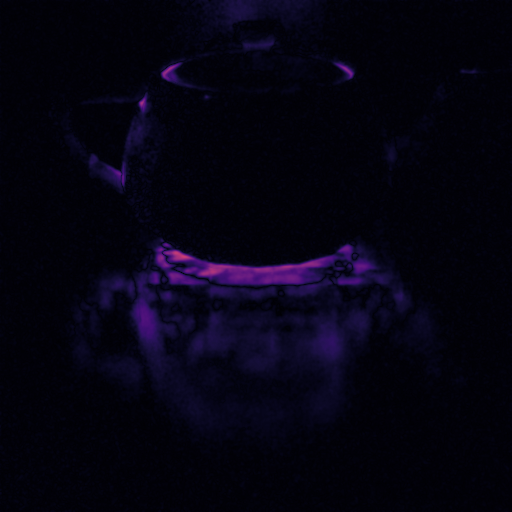} &

  \includegraphics[width=0.08\textwidth]{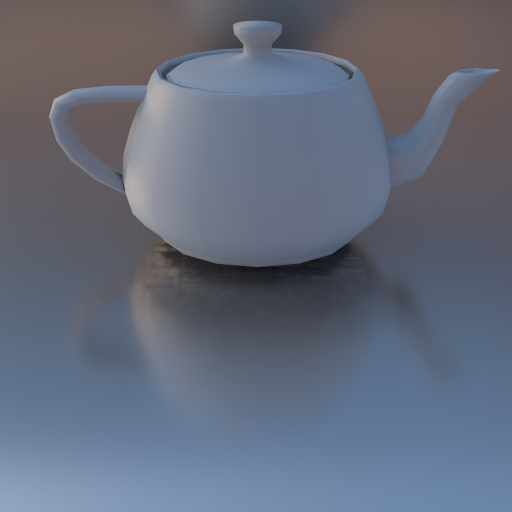} &
  \includegraphics[width=0.08\textwidth]{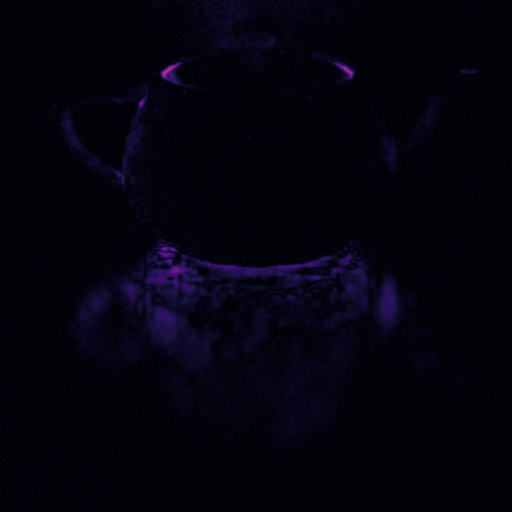} &

  \includegraphics[width=0.08\textwidth]{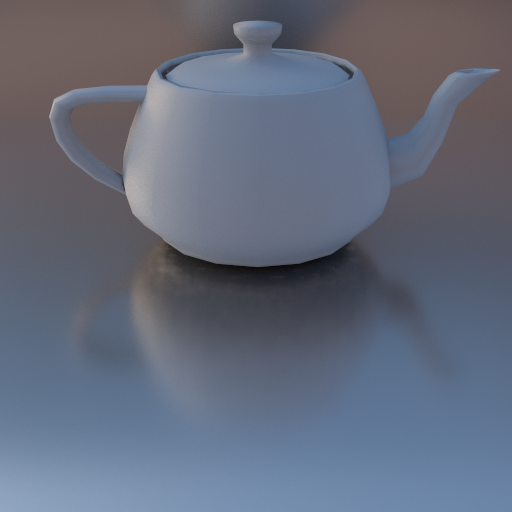} &
  \includegraphics[width=0.08\textwidth]{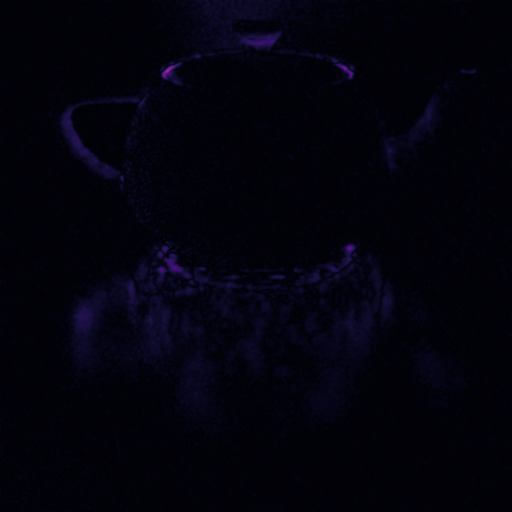} &

  \includegraphics[width=0.08\textwidth]{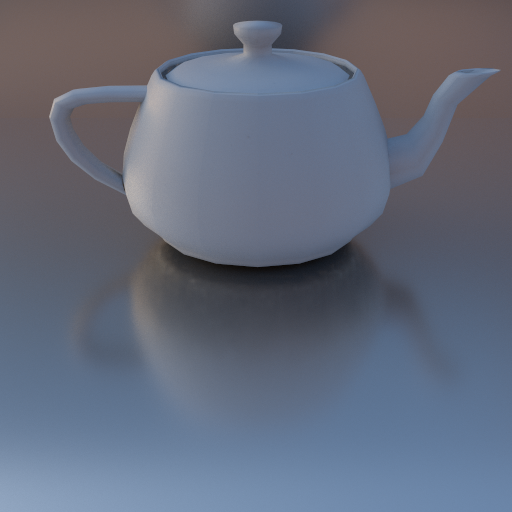} &
  \includegraphics[width=0.08\textwidth]{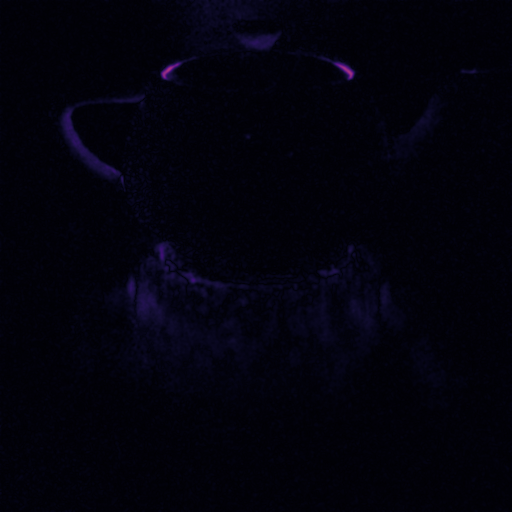} \vspace{-0.205em}\\

  &
  {\footnotesize $0.040$} &

  &
  {\footnotesize $0.034$} &

  &
  {\footnotesize $0.029$} &

  &
  {\footnotesize $0.024$} &

  &
  {\footnotesize $0.022$} &

  &
  {\footnotesize $0.020$} \vspace{-0.275em}\\
  \end{tabular}
  \caption{Comparison on different voxel resolutions, $V$. The four images for each are reconstructed shading normals from LSNIF, a visualization of voxelized geometry, a rendered image using LSNIF, and its FLIP error.}
  \label{fig:results:voxelres}
\end{figure*}
\begin{figure*}[tb]
  \setlength{\fboxrule}{0.002\linewidth}
  \setlength{\fboxsep}{0\linewidth}
        \centering
        \begin{tabular}{c@{\hspace{0.002\linewidth}}c@{\hspace{0.002\linewidth}}c@{\hspace{0.002\linewidth}}c@{\hspace{0.002\linewidth}}c@{\hspace{0.002\linewidth}}c@{\hspace{0.002\linewidth}}c@{\hspace{0.002\linewidth}}c@{\hspace{0.002\linewidth}}c@{\hspace{0.002\linewidth}}c@{\hspace{0.002\linewidth}}}
            \raisebox{-0.68\height}[0pt][0pt]{\includegraphics[width=0.35\textwidth]{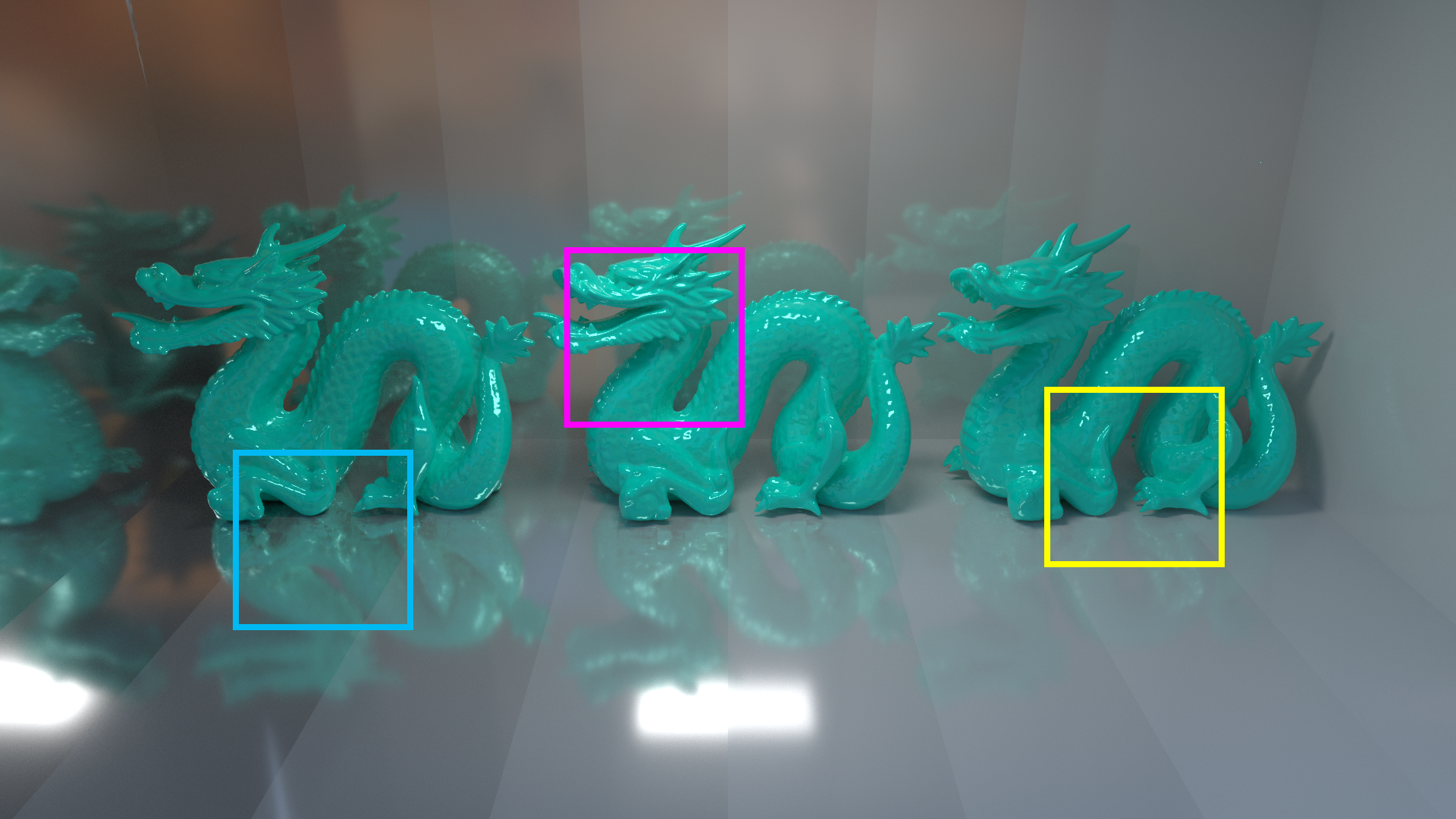}} &
            \fcolorbox{cyan}{cyan}{\includegraphics[width=0.061\textwidth]{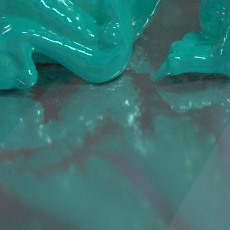}} &
            \fcolorbox{cyan}{cyan}{\includegraphics[width=0.061\textwidth]{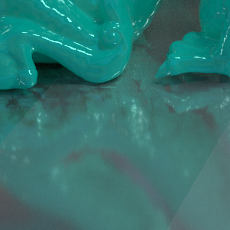}} &
            \fcolorbox{cyan}{cyan}{\includegraphics[width=0.061\textwidth]{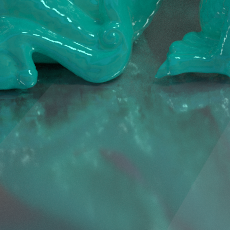}} &
            \fcolorbox{cyan}{cyan}{\includegraphics[width=0.061\textwidth]{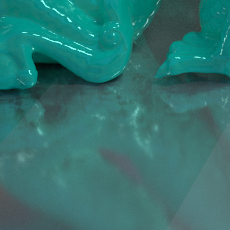}} &
            \fcolorbox{cyan}{cyan}{\includegraphics[width=0.061\textwidth]{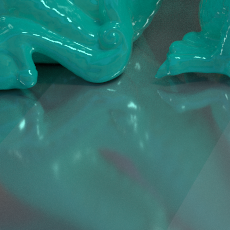}} &
            \fcolorbox{cyan}{cyan}{\includegraphics[width=0.061\textwidth]{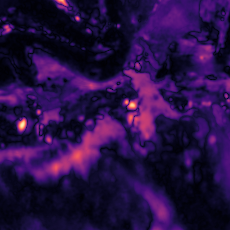}} &
            \fcolorbox{cyan}{cyan}{\includegraphics[width=0.061\textwidth]{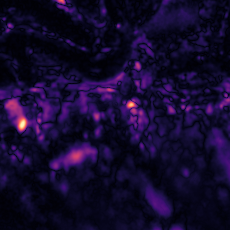}} &
            \fcolorbox{cyan}{cyan}{\includegraphics[width=0.061\textwidth]{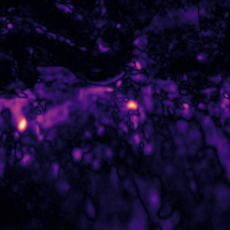}} &
            \fcolorbox{cyan}{cyan}{\includegraphics[width=0.061\textwidth]{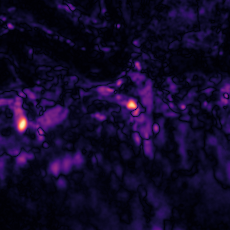}} \vspace{-0.175em}\\

            &
            \fcolorbox{mypink}{mypink}{\includegraphics[width=0.061\textwidth]{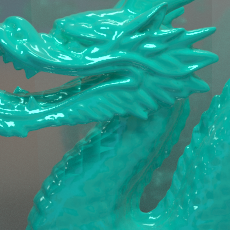}} &
            \fcolorbox{mypink}{mypink}{\includegraphics[width=0.061\textwidth]{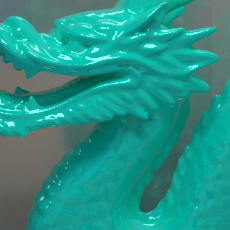}} &
            \fcolorbox{mypink}{mypink}{\includegraphics[width=0.061\textwidth]{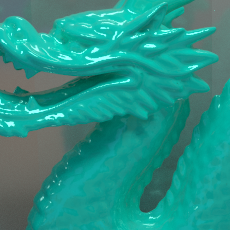}} &
            \fcolorbox{mypink}{mypink}{\includegraphics[width=0.061\textwidth]{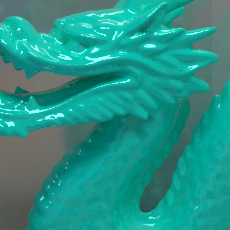}} &
            \fcolorbox{mypink}{mypink}{\includegraphics[width=0.061\textwidth]{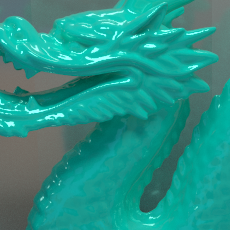}} &
            \fcolorbox{mypink}{mypink}{\includegraphics[width=0.061\textwidth]{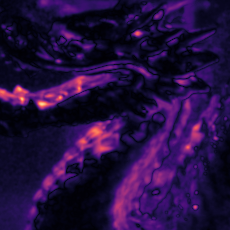}} &
            \fcolorbox{mypink}{mypink}{\includegraphics[width=0.061\textwidth]{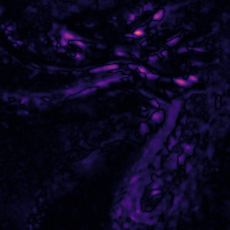}} &
            \fcolorbox{mypink}{mypink}{\includegraphics[width=0.061\textwidth]{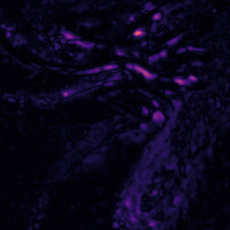}} &
            \fcolorbox{mypink}{mypink}{\includegraphics[width=0.061\textwidth]{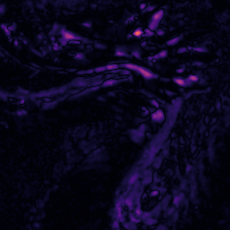}} \vspace{-0.175em}\\

            &
            \fcolorbox{myyellow}{myyellow}{\includegraphics[width=0.061\textwidth]{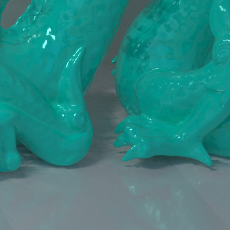}} &
            \fcolorbox{myyellow}{myyellow}{\includegraphics[width=0.061\textwidth]{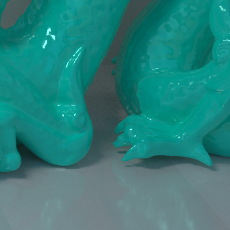}} &
            \fcolorbox{myyellow}{myyellow}{\includegraphics[width=0.061\textwidth]{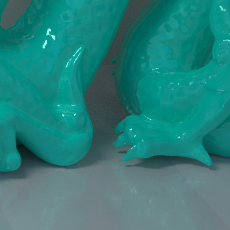}} &
            \fcolorbox{myyellow}{myyellow}{\includegraphics[width=0.061\textwidth]{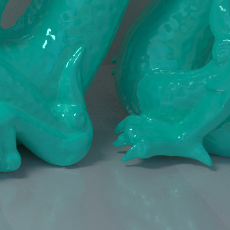}} &
            \fcolorbox{myyellow}{myyellow}{\includegraphics[width=0.061\textwidth]{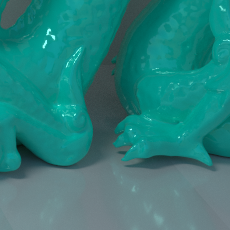}} &
            \fcolorbox{myyellow}{myyellow}{\includegraphics[width=0.061\textwidth]{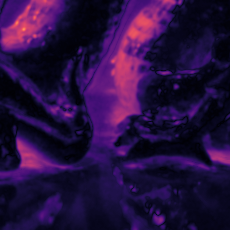}} &
            \fcolorbox{myyellow}{myyellow}{\includegraphics[width=0.061\textwidth]{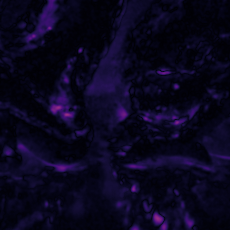}} &
            \fcolorbox{myyellow}{myyellow}{\includegraphics[width=0.061\textwidth]{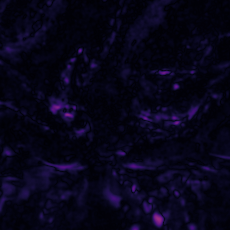}} &
            \fcolorbox{myyellow}{myyellow}{\includegraphics[width=0.061\textwidth]{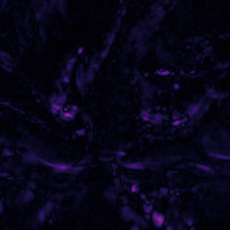}} \vspace{-0.175em}\\

            &
            {\footnotesize $H = 4$} &
            {\footnotesize $H = 8$} &
            {\footnotesize $H = 12$} &
            {\footnotesize $H = 18$} &
            {\footnotesize Ref.} &
            {\footnotesize $H = 4$} &
            {\footnotesize $H = 8$} &
            {\footnotesize $H = 12$} &
            {\footnotesize $H = 18$} \vspace{-0.175em}\\

            &
            &
            &
            &
            &
            &
            {\footnotesize $0.061$} &
            {\footnotesize $0.039$} &
            {\footnotesize $0.035$} &
            {\footnotesize $0.034$} \\
            \raisebox{-0.68\height}[0pt][0pt]{\includegraphics[width=0.35\textwidth]{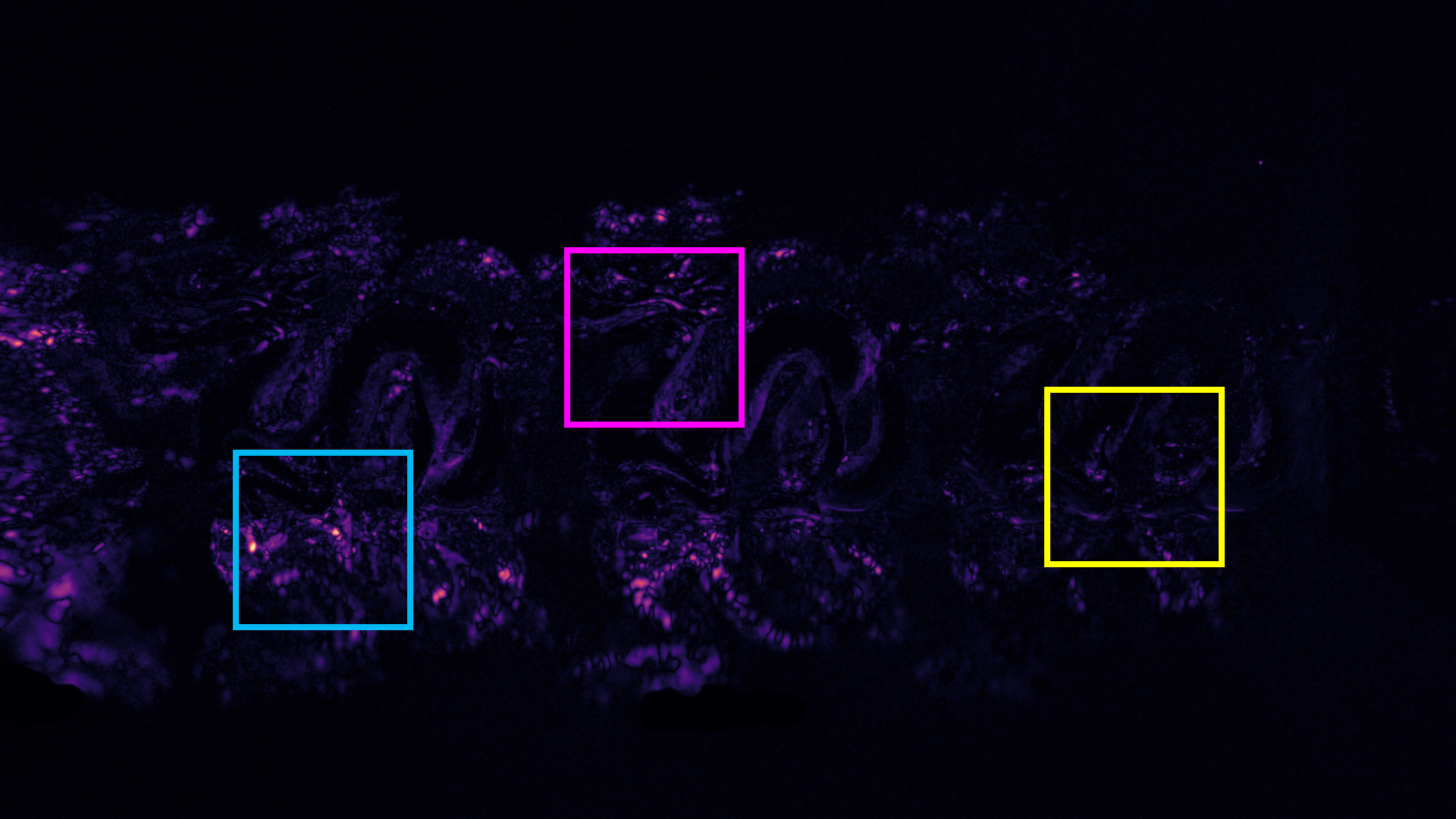}} &
            \fcolorbox{cyan}{cyan}{\includegraphics[width=0.061\textwidth]{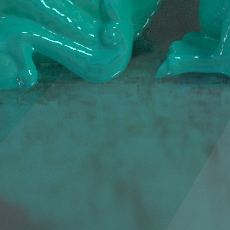}} &
            \fcolorbox{cyan}{cyan}{\includegraphics[width=0.061\textwidth]{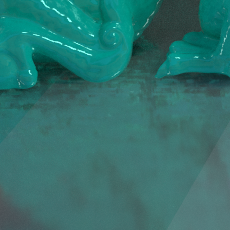}} &
            \fcolorbox{cyan}{cyan}{\includegraphics[width=0.061\textwidth]{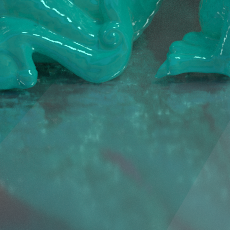}} &
            \fcolorbox{cyan}{cyan}{\includegraphics[width=0.061\textwidth]{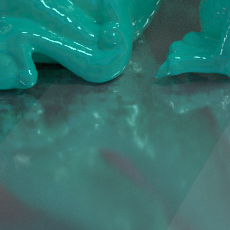}} &
            \fcolorbox{cyan}{cyan}{\includegraphics[width=0.061\textwidth]{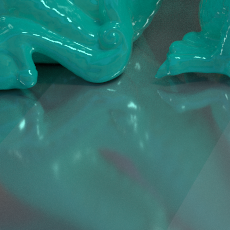}} &
            \fcolorbox{cyan}{cyan}{\includegraphics[width=0.061\textwidth]{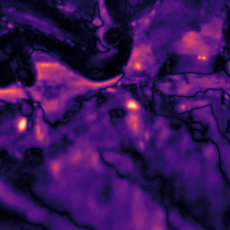}} &
            \fcolorbox{cyan}{cyan}{\includegraphics[width=0.061\textwidth]{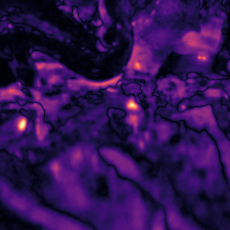}} &
            \fcolorbox{cyan}{cyan}{\includegraphics[width=0.061\textwidth]{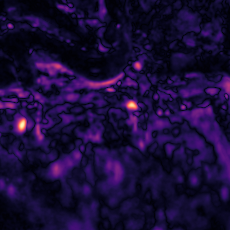}} &
            \fcolorbox{cyan}{cyan}{\includegraphics[width=0.061\textwidth]{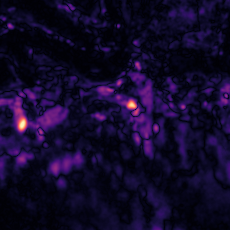}} \vspace{-0.175em}\\

            &
            \fcolorbox{mypink}{mypink}{\includegraphics[width=0.061\textwidth]{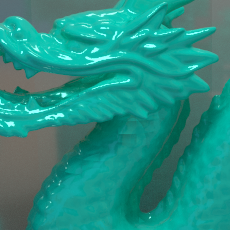}} &
            \fcolorbox{mypink}{mypink}{\includegraphics[width=0.061\textwidth]{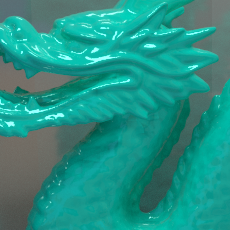}} &
            \fcolorbox{mypink}{mypink}{\includegraphics[width=0.061\textwidth]{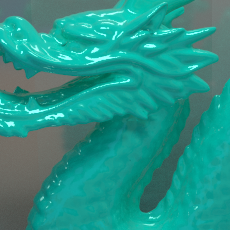}} &
            \fcolorbox{mypink}{mypink}{\includegraphics[width=0.061\textwidth]{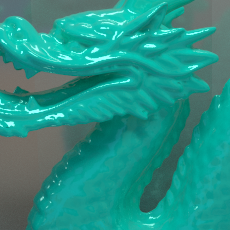}} &
            \fcolorbox{mypink}{mypink}{\includegraphics[width=0.061\textwidth]{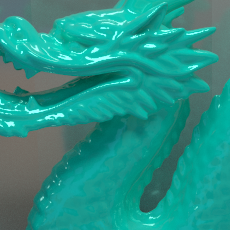}} &
            \fcolorbox{mypink}{mypink}{\includegraphics[width=0.061\textwidth]{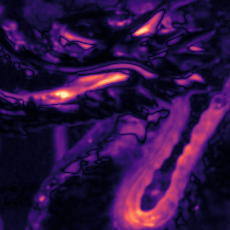}} &
            \fcolorbox{mypink}{mypink}{\includegraphics[width=0.061\textwidth]{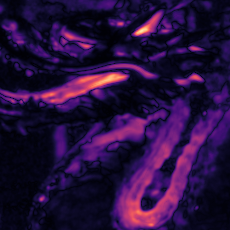}} &
            \fcolorbox{mypink}{mypink}{\includegraphics[width=0.061\textwidth]{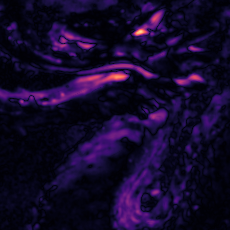}} &
            \fcolorbox{mypink}{mypink}{\includegraphics[width=0.061\textwidth]{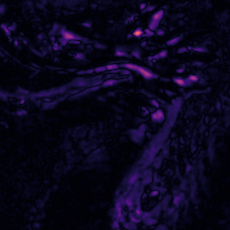}} \vspace{-0.175em}\\

            &
            \fcolorbox{myyellow}{myyellow}{\includegraphics[width=0.061\textwidth]{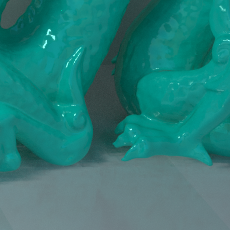}} &
            \fcolorbox{myyellow}{myyellow}{\includegraphics[width=0.061\textwidth]{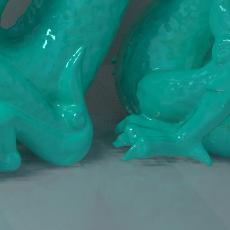}} &
            \fcolorbox{myyellow}{myyellow}{\includegraphics[width=0.061\textwidth]{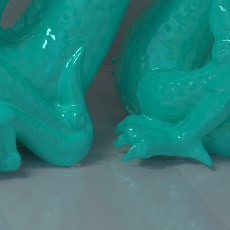}} &
            \fcolorbox{myyellow}{myyellow}{\includegraphics[width=0.061\textwidth]{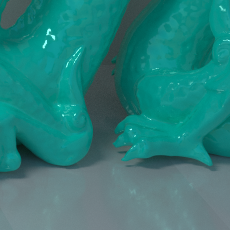}} &
            \fcolorbox{myyellow}{myyellow}{\includegraphics[width=0.061\textwidth]{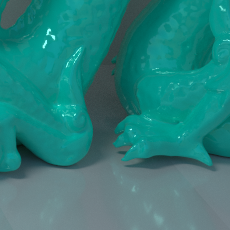}} &
            \fcolorbox{myyellow}{myyellow}{\includegraphics[width=0.061\textwidth]{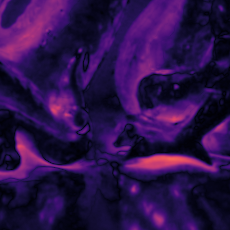}} &
            \fcolorbox{myyellow}{myyellow}{\includegraphics[width=0.061\textwidth]{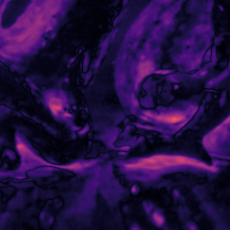}} &
            \fcolorbox{myyellow}{myyellow}{\includegraphics[width=0.061\textwidth]{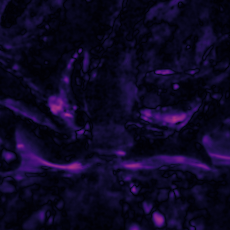}} &
            \fcolorbox{myyellow}{myyellow}{\includegraphics[width=0.061\textwidth]{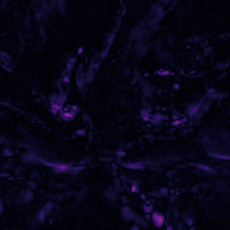}} \vspace{-0.175em}\\

            &
            {\footnotesize $M = 2^8$} &
            {\footnotesize $M = 2^{11}$} &
            {\footnotesize $M = 2^{14}$} &
            {\footnotesize $M = 2^{17}$} &
            {\footnotesize Ref.} &
            {\footnotesize $M = 2^8$} &
            {\footnotesize $M = 2^{11}$} &
            {\footnotesize $M = 2^{14}$} &
            {\footnotesize $M = 2^{17}$} \vspace{-0.175em}\\

            &
            {\footnotesize $3$ KB} &
            {\footnotesize $24$ KB} &
            {\footnotesize $192$ KB} &
            {\footnotesize $1.5$ MB} &
            &
            {\footnotesize $0.085$} &
            {\footnotesize $0.061$} &
            {\footnotesize $0.047$} &
            {\footnotesize $0.034$} \\
        \end{tabular}
        \caption{ Comparison of different upper bounds of the number of intersection points, $H$, and different hash-map sizes, $M$. The top-left image is rendered using LSNIF with $H = 18$ and $M = 2^{17}$ and the bottom-left image is its FLIP error. All images are rendered with 10k spp after 2,000 training steps. }
        \label{fig:results:hitcap}
\end{figure*}

We demonstrate the effectiveness of LSNIF by rendering various scenes with $10$k spps with $1920 \times 1080$ screen resolution on an AMD Radeon\textsuperscript{\texttrademark} RX 7900 XTX GPU.
The rendered images are compared with the reference images generated by the BVH-based path tracer using the FLIP error metric, where a lower value indicates a better reconstruction quality~\cite{Andersson2020}.

\subsection{Parameter Study} \label{sec:parameter}
There are a few parameters that determine the accuracy of LSNIF. In this section, we show how these parameters affect the rendering results.
The first parameter that has a substantial impact on the reconstruction quality is the voxel resolution, $V$.
Higher $V$ makes the voxel representation closer to the underlying geometry, as shown in Fig.~\ref{fig:results:voxelres}.
On the other hand, when lower $V$ is used, LSNIF gets closer to NIF~\cite{NIF}.
LSNIF can carve out the teapot shape even when $2^3$ resolution is used, but it results in obvious errors in the geometric representation that is visible in reflections. This is not exactly the same reproduction of NIF but it is close enough to be compared.
This shows the results that we would get if we trained NIF offline and used it in the rendering. 
The higher $V$ we use, the smaller the error becomes.
In the example of Fig.~\ref{fig:results:voxelres}, artifacts are not visible at $32^3$ resolution.
Since the voxel data contains only binary occupancies, its memory footprint is negligible. It increases from $4$ KB to $32$ KB when the voxel resolution is changed from $32^3$ to $64^3$.

Another parameter, $H$, which controls the image quality is how many points we collect for the NN inputs.
Smaller $H$ works fine for simple geometries.
However, it fails when a ray passes through several voxels, coming close to the underlying geometry but without the actual hits, as illustrated with an orange ray in Fig.~\ref{fig:3:flow}.
These errors appear as clear visual artifacts as shown in the top part of Fig.~\ref{fig:results:hitcap}.
When $H$ is small, some parts of the geometry are missing (top row), self-shadows are not cast (middle row), or shadows cast to another geometry are not visible (bottom row).
This is because small $H$ does not have enough capacity to capture all the hits to the voxels where the ray hits the underlying geometry, thus it fails to reconstruct the geometry.
Note that $H$ depends on $V$. As we use higher $V$, we need to increase $H$ accordingly.
We experimentally choose $V = 32^3$ and $H = 18$, and use these values for all other experiments in this paper.

The major parameter that significantly impacts the memory footprint is the hash-map size, $M$, in the sparse grid encoding, which takes up a large part of LSNIF's memory.
The bottom part of Fig.~\ref{fig:results:hitcap} also compares the memory footprint with various $M$ and how they affect the reconstruction quality.
Lower $M$ causes more hash collisions, thus more points are mapped to the same memory. This results in clear visual artifacts in reflections and shadows.
We find $M = 2^{17}$ is a good balance between quality and memory overhead.

\begin{figure*}[tb]
  \setlength{\fboxrule}{0.002\linewidth}
  \setlength{\fboxsep}{0\linewidth}
        \centering
        \begin{tabular}{c@{\hspace{0.002\linewidth}}c@{\hspace{0.002\linewidth}}c@{\hspace{0.002\linewidth}}c@{\hspace{0.002\linewidth}}c}
            &
            {\footnotesize $T = 100$} &
            {\footnotesize $T = 700$} &
            {\footnotesize $T = 2000$} &
            \\

            \raisebox{-0.68\height}[0pt][0pt]{\includegraphics[width=0.39\textwidth]{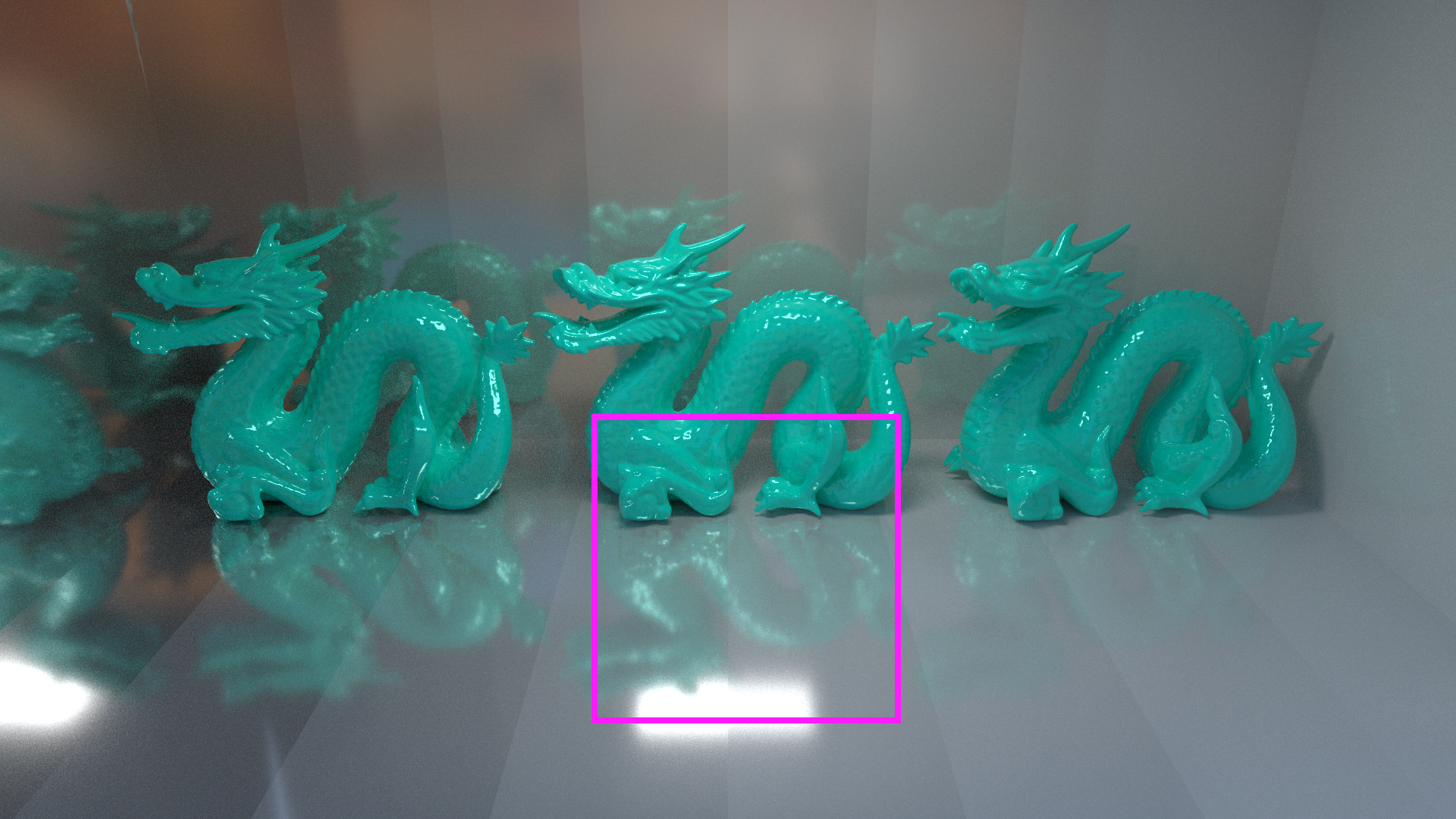}} &
            \includegraphics[width=0.07\textwidth]{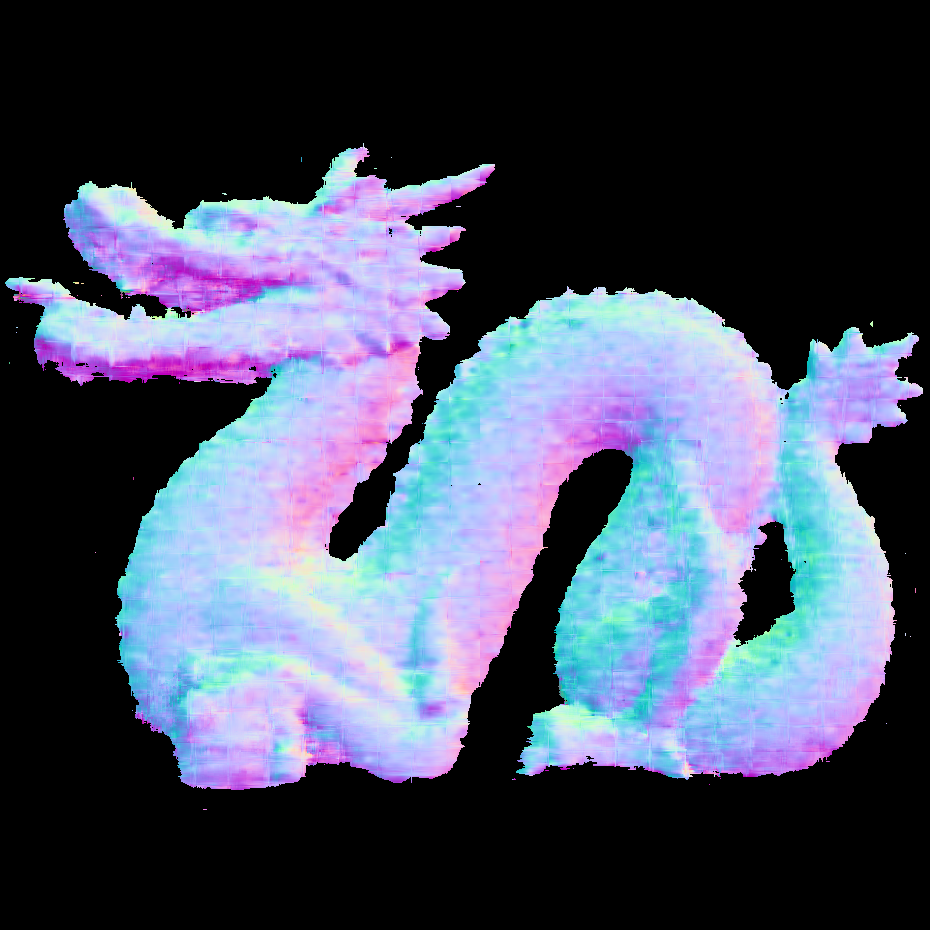} &
            \includegraphics[width=0.07\textwidth]{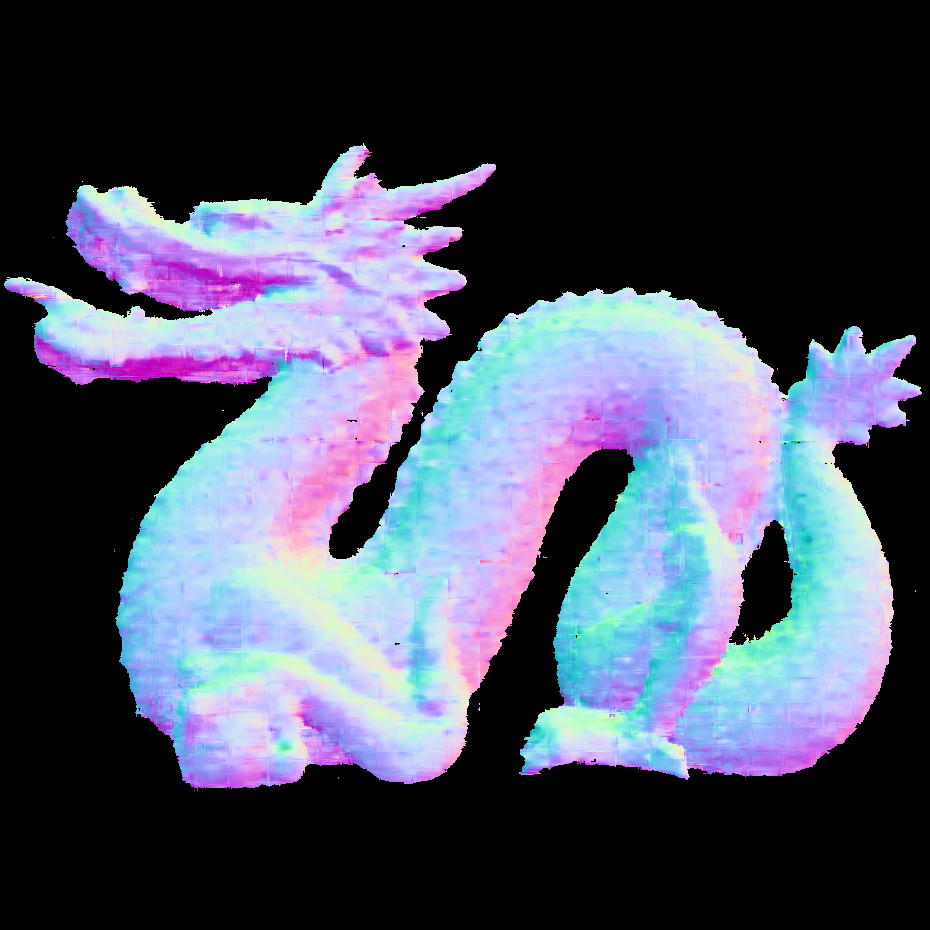} &
            \includegraphics[width=0.07\textwidth]{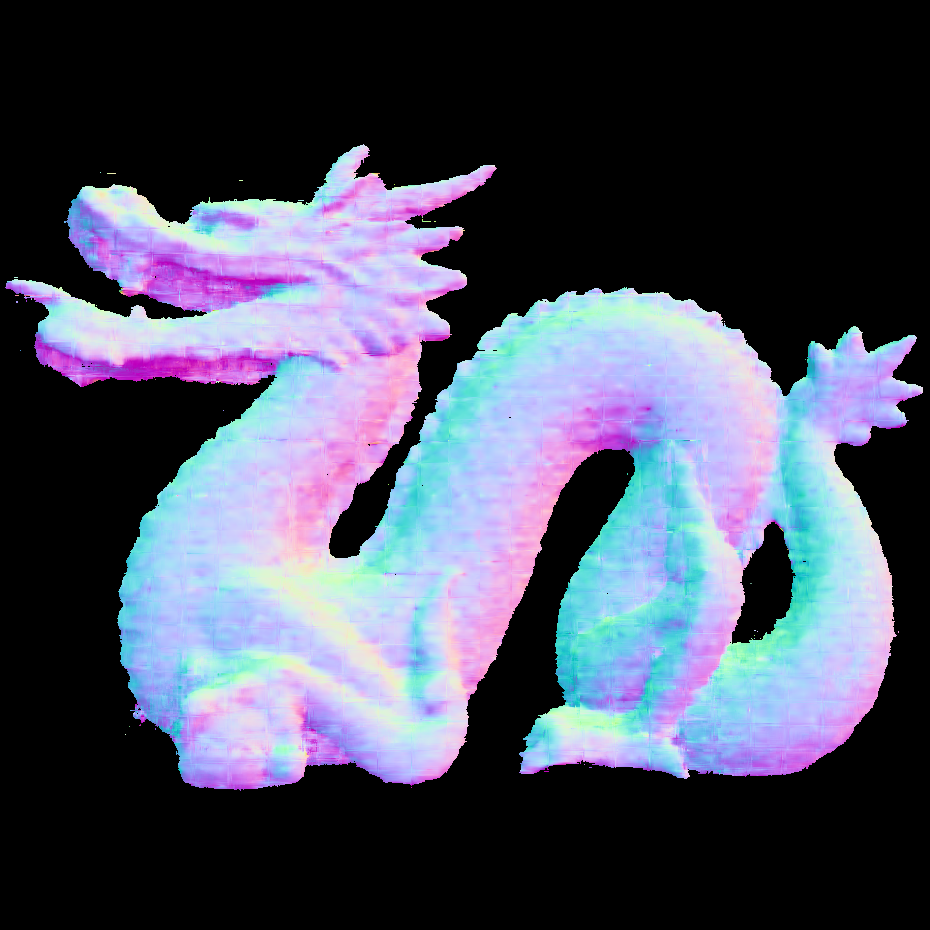} &
            \raisebox{-0.63\height}[0pt][0pt]{\includegraphics[width=0.38\textwidth]{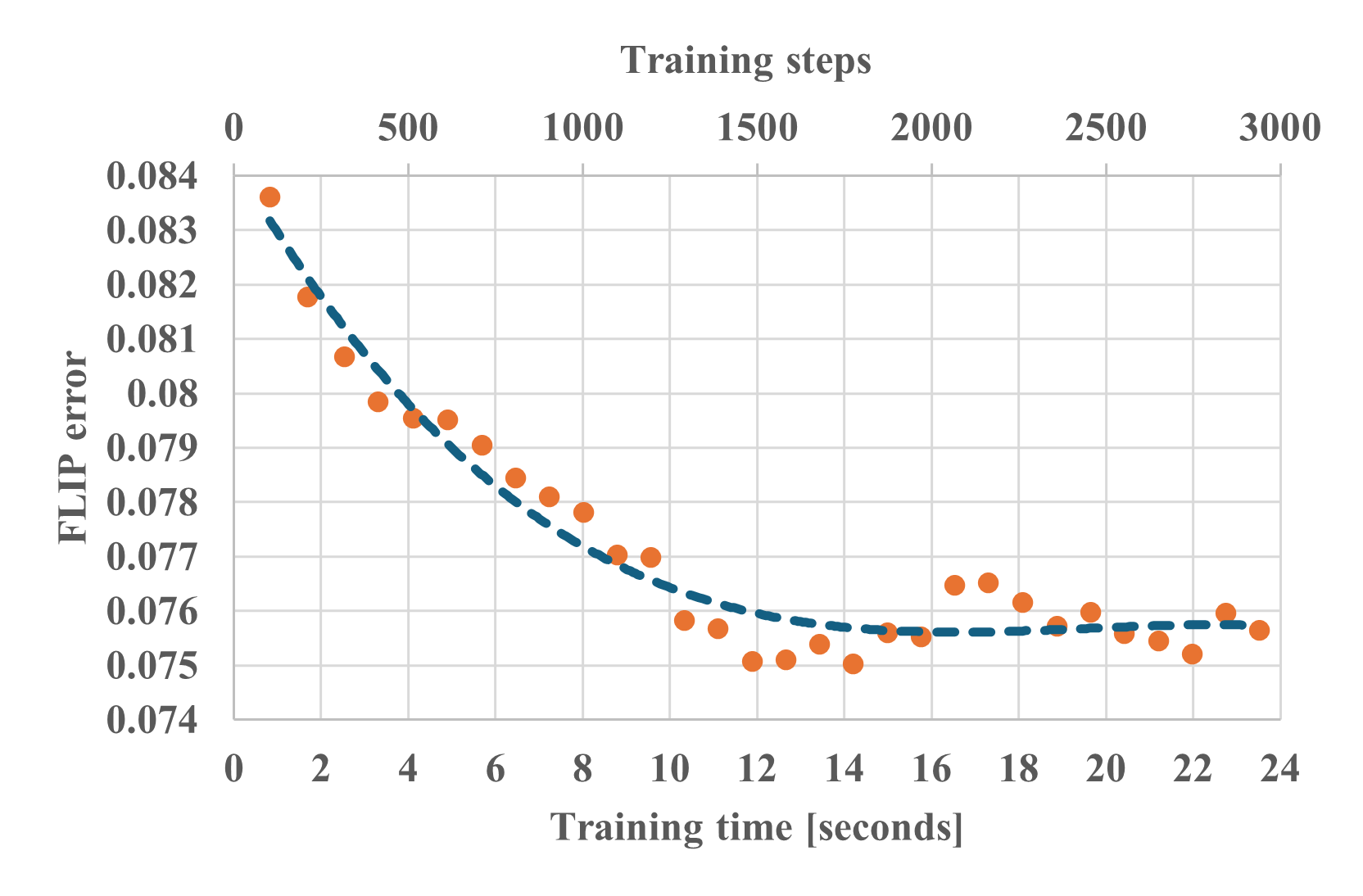}} \vspace{-0.175em} \\

            &
            \fcolorbox{mypink}{mypink}{\includegraphics[width=0.068\textwidth]{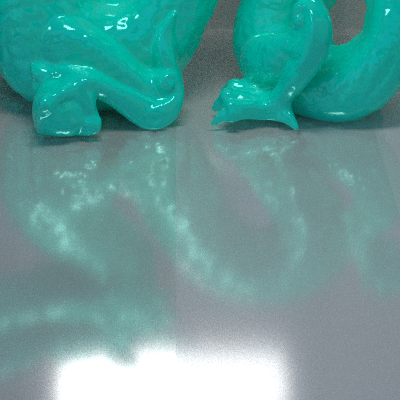}} &
            \fcolorbox{mypink}{mypink}{\includegraphics[width=0.068\textwidth]{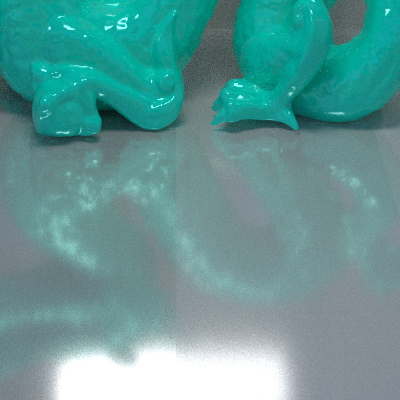}} &
            \fcolorbox{mypink}{mypink}{\includegraphics[width=0.068\textwidth]{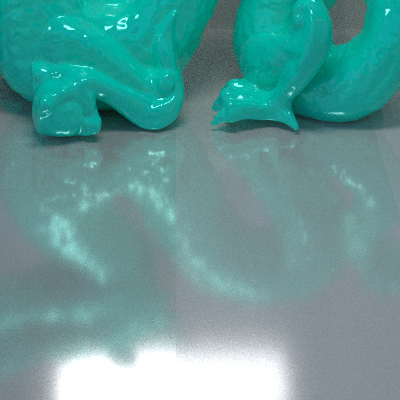}} &
            \vspace{-0.175em}\\

            &
            \fcolorbox{mypink}{mypink}{\includegraphics[width=0.068\textwidth]{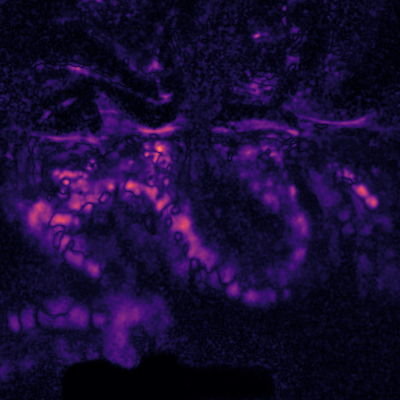}} &
            \fcolorbox{mypink}{mypink}{\includegraphics[width=0.068\textwidth]{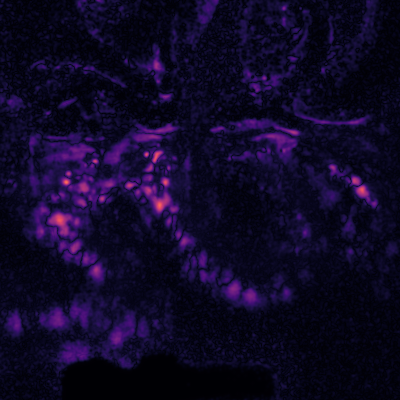}} &
            \fcolorbox{mypink}{mypink}{\includegraphics[width=0.068\textwidth]{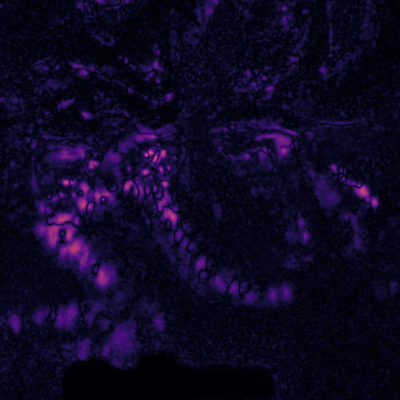}} &
            \vspace{-0.175em}\\

            &
            {\footnotesize $0.060$} &
            {\footnotesize $0.049$} &
            {\footnotesize $0.046$} &
            \vspace{-0.225em}\\
        \end{tabular}
        \caption{ Comparison of reconstruction quality with different training steps, $T$. Top-middle images are reconstructed first-hit shading normals from LSNIF. Other close-up images show rendered images using LSNIF, and FLIP errors for different training timings. The left image is rendered with 2,000 training steps, which shows enough quality and takes only about 16 seconds for training. }
        \label{fig:results:trainingtime}
\end{figure*}

\begin{figure*}
  \setlength{\fboxrule}{0.002\linewidth}
  \setlength{\fboxsep}{0\linewidth}
  \setlength{\tabcolsep}{0.002\linewidth}
    \begin{tabular}{c@{\hspace{0.001\linewidth}}c@{\hspace{0.001\linewidth}}c}
    \textsc{Junkshop}&
    \textsc{Hairballs}&
    \textsc{Botanic Cornell Box Glossy}  \vspace{-0.175em}\\
    \includegraphics[width=0.32\textwidth]{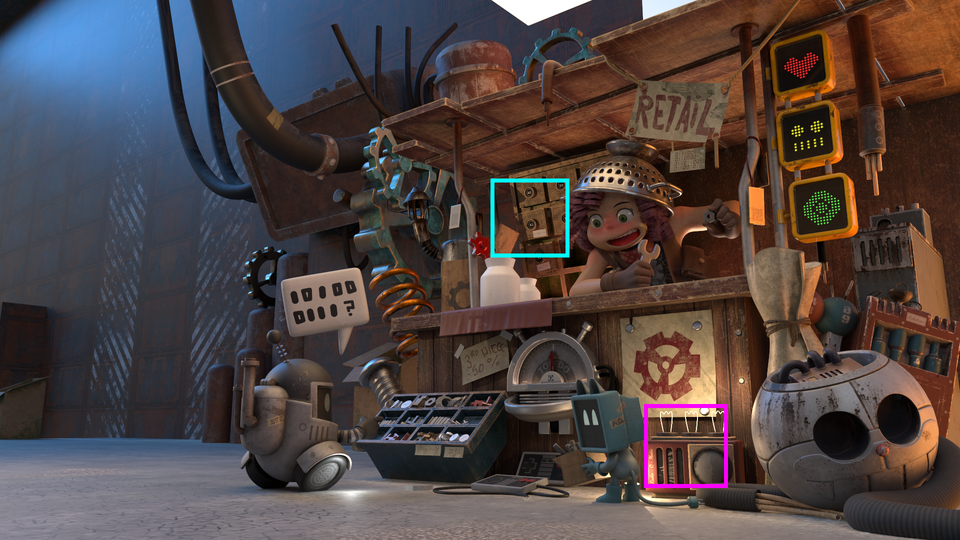} &
    \includegraphics[width=0.32\textwidth]{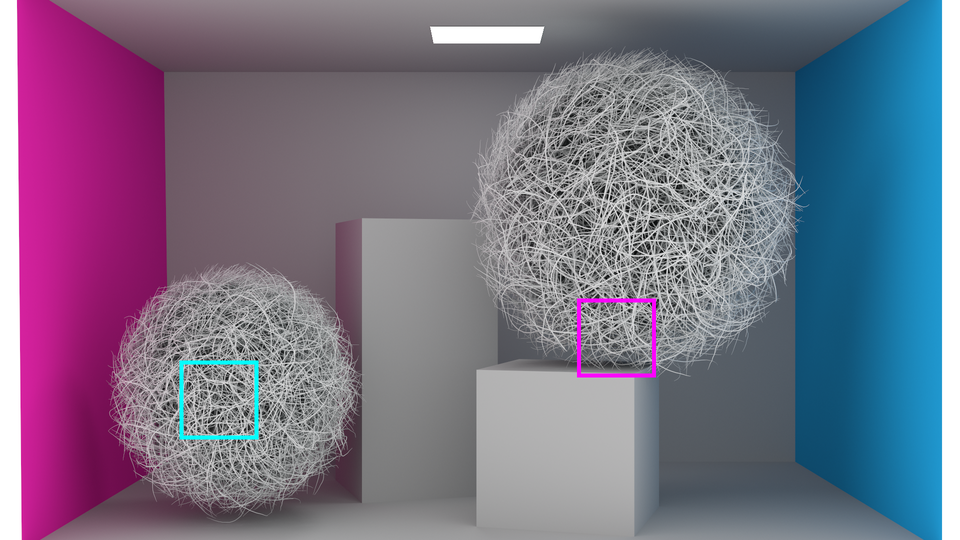} &
    \includegraphics[width=0.32\textwidth]{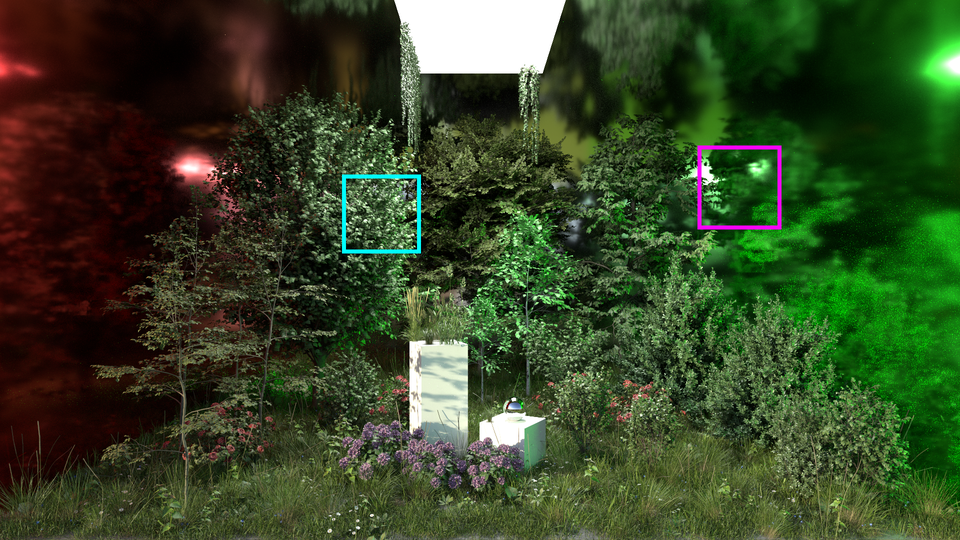} \vspace{-0.275em}\\
    \begin{tabular}{c@{\hspace{0.002\linewidth}}c}
        \includegraphics[width=0.159\textwidth]{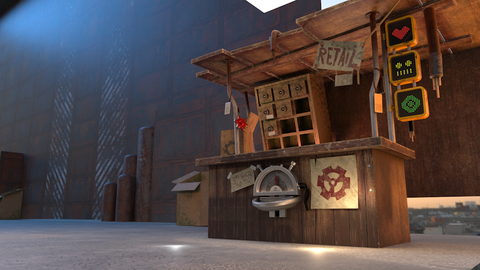} & 
        \includegraphics[width=0.159\textwidth]{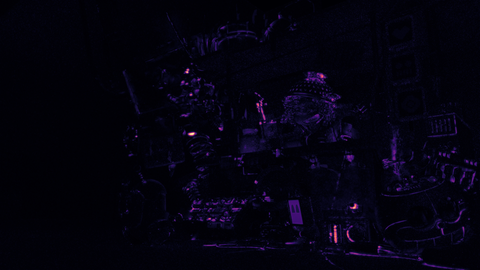} 
    \end{tabular}
    &
    \begin{tabular}{c@{\hspace{0.002\linewidth}}c}    
        \includegraphics[width=0.159\textwidth]{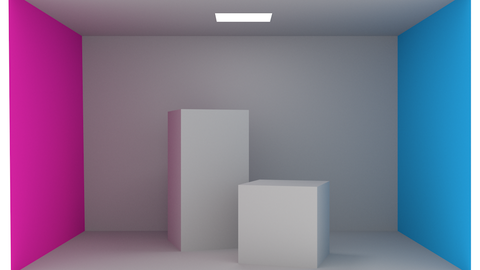} & 
        \includegraphics[width=0.159\textwidth]{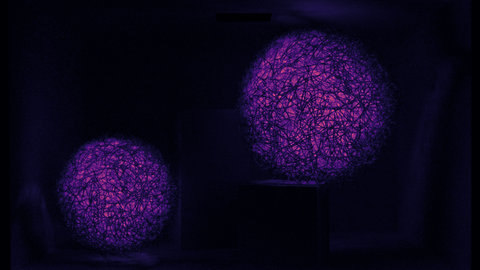}
    \end{tabular}
    & 
    \begin{tabular}{c@{\hspace{0.002\linewidth}}c}
        \includegraphics[width=0.159\textwidth]{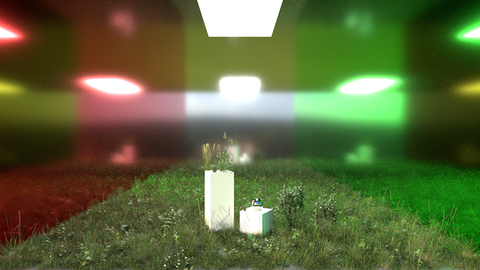} & 
        \includegraphics[width=0.159\textwidth]{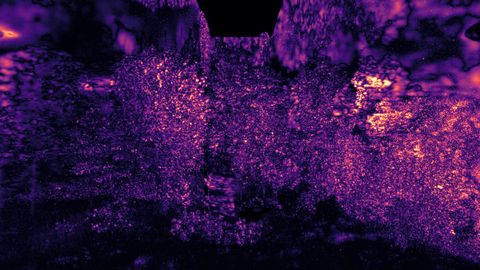}
    \end{tabular}
    \vspace{-0.275em}\\
    \hspace*{0.01em}
    \begin{tabular}{c@{\hspace{0.001\linewidth}}c@{\hspace{0.001\linewidth}}c}
        \fcolorbox{cyan}{cyan}{\includegraphics[width=0.102\textwidth]{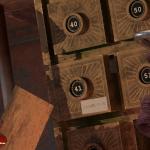}} & 
        \fcolorbox{cyan}{cyan}{\includegraphics[width=0.102\textwidth]{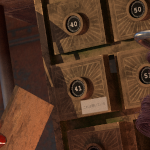}} & 
        \fcolorbox{cyan}{cyan}{\includegraphics[width=0.102\textwidth]{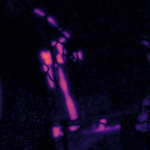}} \vspace{-0.275em}\\
        \fcolorbox{mypink}{mypink}{\includegraphics[width=0.102\textwidth]{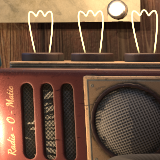}} & 
        \fcolorbox{mypink}{mypink}{\includegraphics[width=0.102\textwidth]{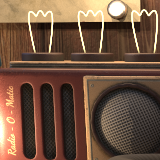}} & 
        \fcolorbox{mypink}{mypink}{\includegraphics[width=0.102\textwidth]{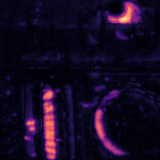}} \vspace{-0.275em}\\
        {\footnotesize LSNIF} & 
        {\footnotesize Reference} & 
        {\footnotesize FLIP: $0.037$}  
    \end{tabular}
    &
    \hspace*{0.01em}
    \begin{tabular}{c@{\hspace{0.001\linewidth}}c@{\hspace{0.001\linewidth}}c}
        \fcolorbox{cyan}{cyan}{\includegraphics[width=0.102\textwidth]{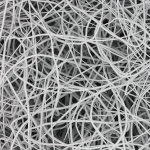}} & 
        \fcolorbox{cyan}{cyan}{\includegraphics[width=0.102\textwidth]{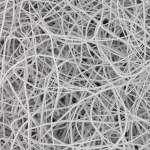}} & 
        \fcolorbox{cyan}{cyan}{\includegraphics[width=0.102\textwidth]{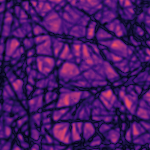}} \vspace{-0.275em}\\
        \fcolorbox{mypink}{mypink}{\includegraphics[width=0.102\textwidth]{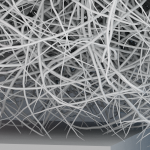}} & 
        \fcolorbox{mypink}{mypink}{\includegraphics[width=0.102\textwidth]{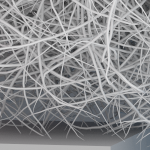}} & 
        \fcolorbox{mypink}{mypink}{\includegraphics[width=0.102\textwidth]{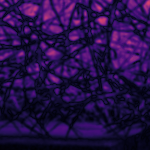}} \vspace{-0.275em}\\
        {\footnotesize LSNIF} & 
        {\footnotesize Reference} & 
        {\footnotesize FLIP: $0.066$}  
    \end{tabular}
    &
    \hspace*{0.01em}
    \begin{tabular}{c@{\hspace{0.001\linewidth}}c@{\hspace{0.001\linewidth}}c}
        \fcolorbox{cyan}{cyan}{\includegraphics[width=0.102\textwidth]{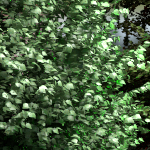}} & 
        \fcolorbox{cyan}{cyan}{\includegraphics[width=0.102\textwidth]{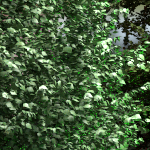}} & 
        \fcolorbox{cyan}{cyan}{\includegraphics[width=0.102\textwidth]{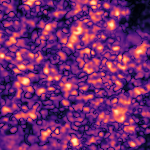}} \vspace{-0.275em}\\
        \fcolorbox{mypink}{mypink}{\includegraphics[width=0.102\textwidth]{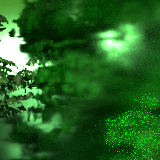}} & 
        \fcolorbox{mypink}{mypink}{\includegraphics[width=0.102\textwidth]{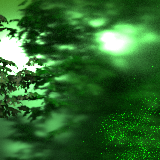}} & 
        \fcolorbox{mypink}{mypink}{\includegraphics[width=0.102\textwidth]{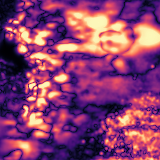}} \vspace{-0.275em}\\
        {\footnotesize LSNIF} & 
        {\footnotesize Reference} & 
        {\footnotesize FLIP: $0.170$}  
    \end{tabular}
    \vspace{-0.275em}
    \end{tabular}
    \caption{First row: rendered images using LSNIF. Second row: rendered image without LSNIF, and FLIP error. The rest are close-ups of rendered images comparing LSNIF and reference. }
    \label{fig:results:morescenes}    
\end{figure*}

\subsection{Training Time}
Although we did not fully optimize our training implementation which transfers data between GPU and CPU, training LSNIF for a single geometry takes only a few tens of seconds, including data generation, i.e., ray casting using a BVH.
Fig.~\ref{fig:results:trainingtime} shows the quality of the rendered images using LSNIF with different training steps, $T$.
In this experiment, to correctly observe the reconstruction errors for different $T$, we computed the FLIP errors for the images rendered using LSNIF for all types of rays including primary rays with $1$k spp. 
The rightmost graph in Fig.~\ref{fig:results:trainingtime} shows the moving averages of FLIP errors over training time. With small $T$, reconstructed shading normals are not accurate, resulting in visible errors in the rendered images.
We find that $T = 2,000$ is sufficient to achieve visually good-quality results, so we use $2,000$ training steps for the rest of the experiments in this paper.
Training LSNIF with $T = 2,000$ takes about 15 seconds for a single geometry on a single GPU.

\begin{figure*}
\centering
  \setlength{\fboxrule}{0.002\linewidth}
  \setlength{\fboxsep}{0\linewidth}
  \setlength{\tabcolsep}{0.002\linewidth}
    \begin{tabular}{c@{\hspace{0.002\linewidth}}c@{\hspace{0.002\linewidth}}c@{\hspace{0.002\linewidth}}c@{\hspace{0.002\linewidth}}c@{\hspace{0.002\linewidth}}c@{\hspace{0.002\linewidth}}c@{\hspace{0.002\linewidth}}c@{\hspace{0.002\linewidth}}c@{\hspace{0.002\linewidth}}}

    \fcolorbox{black}{black}{\includegraphics[width=0.101\textwidth]{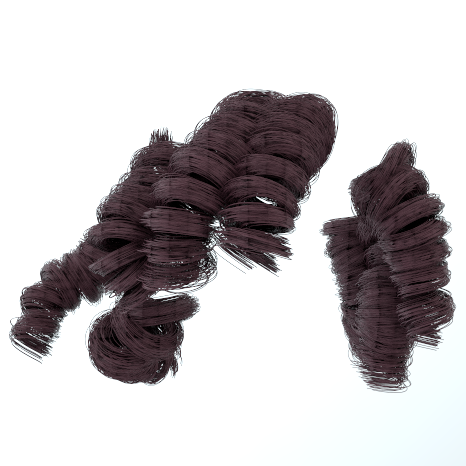}} & 
    \fcolorbox{black}{black}{\includegraphics[width=0.101\textwidth]{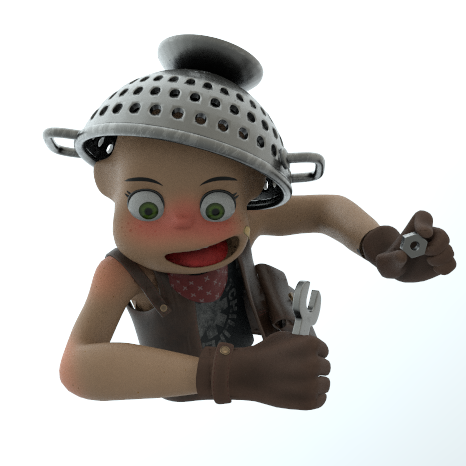}} & 
    \fcolorbox{black}{black}{\includegraphics[width=0.101\textwidth]{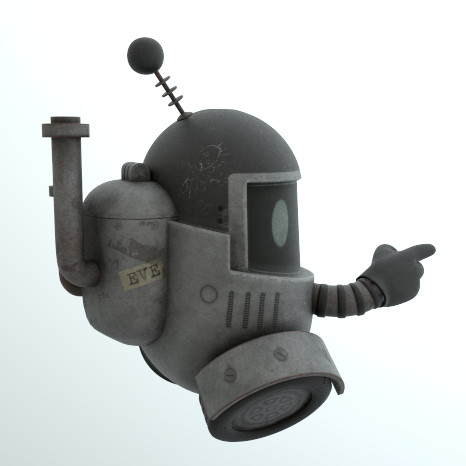}} & 
    
    \fcolorbox{black}{black}{\includegraphics[width=0.101\textwidth]{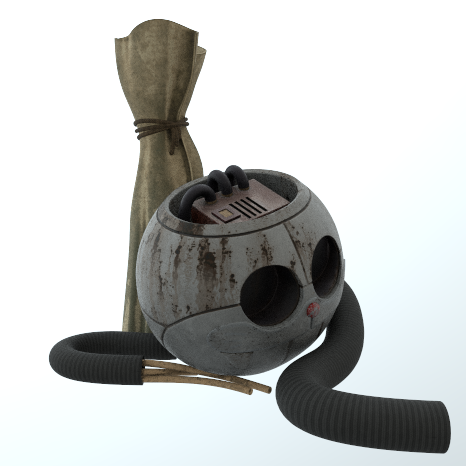}} & 
    \fcolorbox{black}{black}{\includegraphics[width=0.101\textwidth]{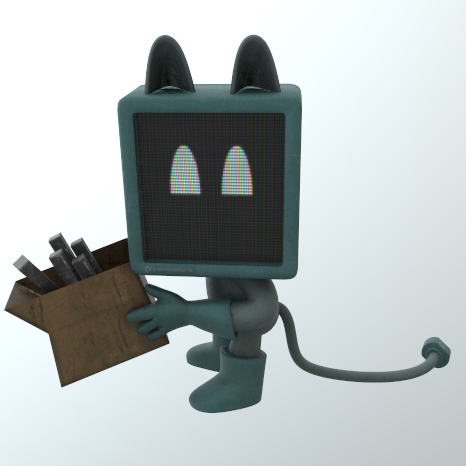}} & 
    \fcolorbox{black}{black}{\includegraphics[width=0.101\textwidth]{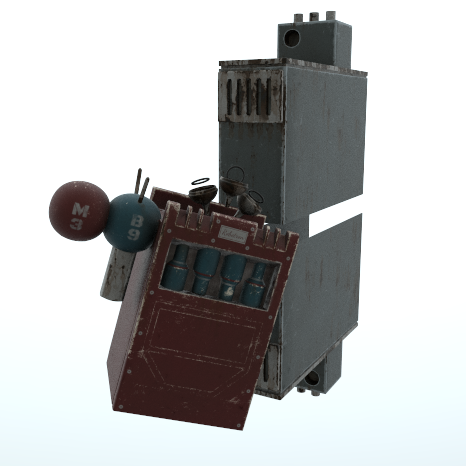}} & 

    \fcolorbox{black}{black}{\includegraphics[width=0.101\textwidth]{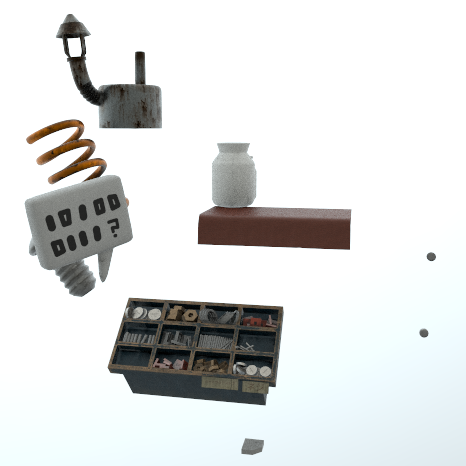}} & 
    \fcolorbox{black}{black}{\includegraphics[width=0.101\textwidth]{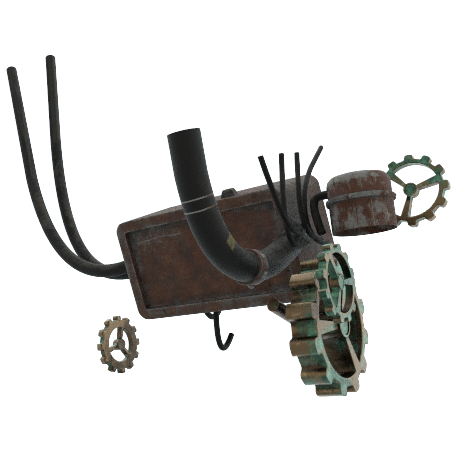}} & 
    \fcolorbox{black}{black}{\includegraphics[width=0.101\textwidth]{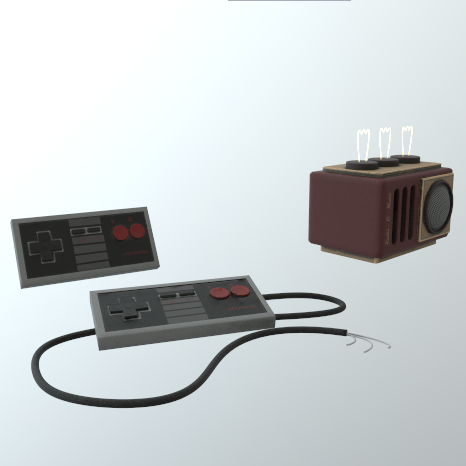}} \\
    \end{tabular} 
    \caption{Rendered images of LSNIF objects in \textsc{Junkshop} scene. All are rendered using LSNIF.}
    \label{fig:junkshopparts}  
\end{figure*}

\begin{figure*}
     \centering
     \begin{subfigure}[b]{0.32\textwidth}
         \centering
         \includegraphics[width=\textwidth]{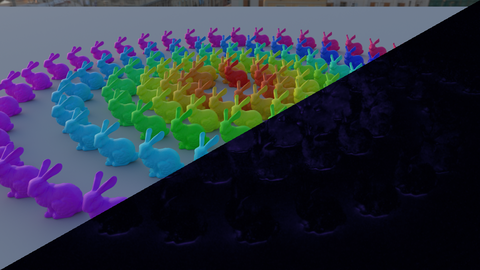}
         \caption{LSNIF: $187.2$ MB, FLIP: $0.028$}
     \end{subfigure}
     \hfill
     \begin{subfigure}[b]{0.32\textwidth}
         \centering
         \includegraphics[width=\textwidth]{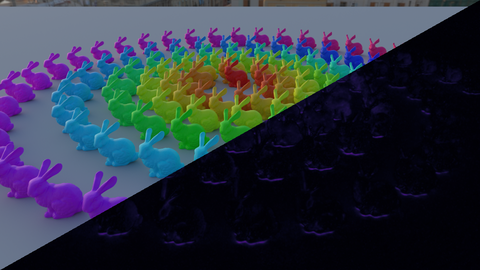}
         \caption{LSNIF: $46.8$ MB, FLIP: $0.029$}
     \end{subfigure}     
     \hfill
     \begin{subfigure}[b]{0.32\textwidth}
         \centering
         \includegraphics[width=\textwidth]{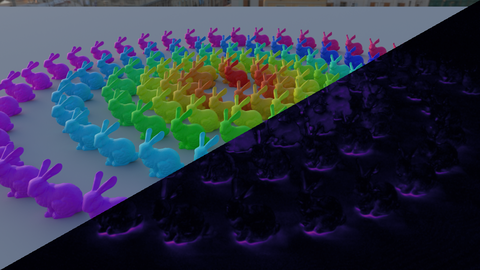}
         \caption{LSNIF: $12.5$ MB, FLIP: $0.044$}
     \end{subfigure}     
        \caption{Comparison of different LSNIF granularity. An LSNIF is created for each bunny (a), 4 bunnies (b), and 15 bunnies (c), respectively. Each image's upper and lower half show the rendered image and FLIP errors. }
        \label{fig:nifgranularity}
\end{figure*}

\subsection{Quality Evaluation}
Fig.~\ref{fig:teaser} and Fig.~\ref{fig:results:morescenes} show various scenes rendered using LSNIF to evaluate the reconstruction quality.
In \textsc{Statues}, all the geometries except for the boxes are represented by LSNIF.
Thus, the BVH used for ray tracing against triangles only contains $36$ triangle primitives when LSNIF is used.
This is a huge complexity reduction to pure ray tracing as there are $18.2$ million triangles in the scene.
Because a single LSNIF object is $1.56$ MB with the parameter discussed in Sec.~\ref{sec:parameter}, LSNIF only consumes $6.24$ MB, while uncompressed BVH for $18.2$ million triangles takes $2.5$ GB for this scene.
Therefore, LSNIF compresses the geometry to only $0.24\%$ with small visual artifacts.
We also tested LSNIF with some highly-detailed geometries, with which BVH ray tracing is computationally expensive, such as foliage and thin geometries in \textsc{Botanic Cornell Box}, \textsc{Botanic Cornell Box Glossy} and \textsc{Hairballs}.
The errors in the rendered images with LSNIF are relatively higher compared to those with simple geometries, mostly due to missing intersections.
However, these results are still visually plausible. Improving these cases is one of the future works.

\begin{figure*}
\centering
  \setlength{\fboxrule}{0.002\linewidth}
  \setlength{\fboxsep}{0\linewidth}
  \setlength{\tabcolsep}{0.002\linewidth}
    \begin{tabular}{c@{\hspace{0.002\linewidth}}c@{\hspace{0.002\linewidth}}c@{\hspace{0.002\linewidth}}c@{\hspace{0.002\linewidth}}c@{\hspace{0.002\linewidth}}}
    \raisebox{0.2em}{\rotatebox[origin=l]{90}{{\scriptsize Lighting Change}}} &
    \includegraphics[width=0.243\textwidth]{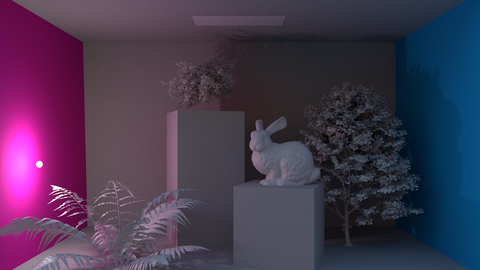} &
    \includegraphics[width=0.243\textwidth]{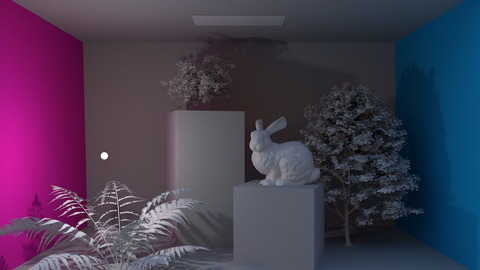}&
    \includegraphics[width=0.243\textwidth]{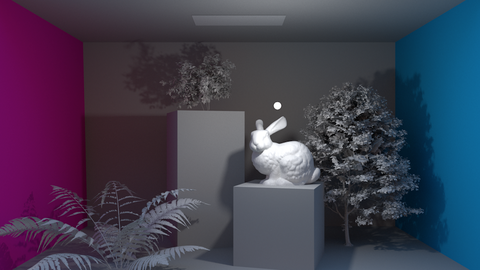} & 
    \includegraphics[width=0.243\textwidth]{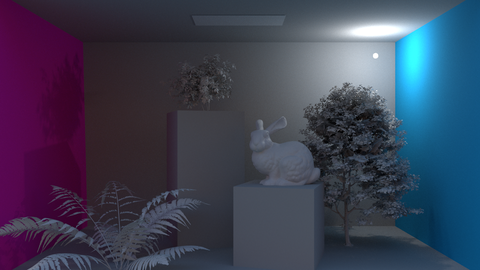} \vspace{-0.175em}\\
    \raisebox{-0.12em}{\rotatebox[origin=l]{90}{{\scriptsize Transform Change}}} &
    \includegraphics[width=0.243\textwidth]{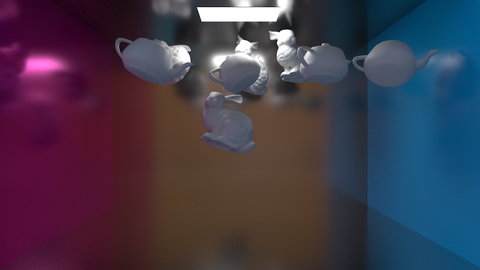} &
    \includegraphics[width=0.243\textwidth]{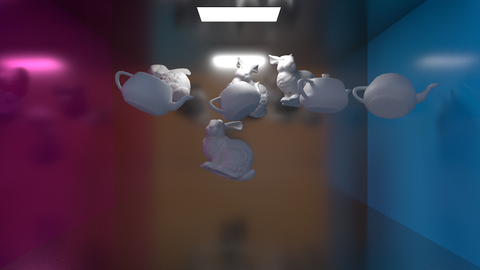}&
    \includegraphics[width=0.243\textwidth]{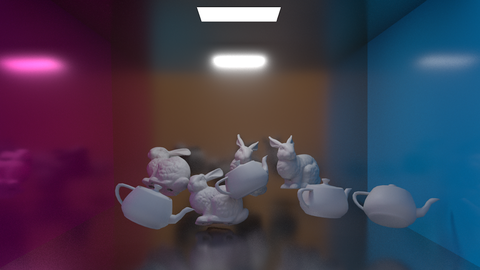} & 
    \includegraphics[width=0.243\textwidth]{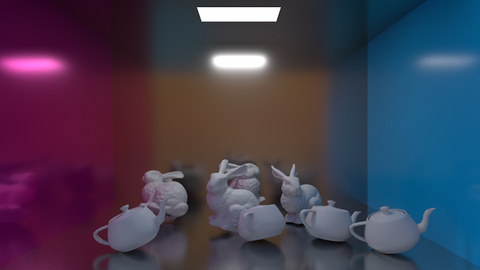} \vspace{-0.175em}\\
    \end{tabular}
    \caption{Frames from an animation sequence where light and transformation of objects are changing. All objects except for the Cornell box itself and the spherical light are represented as LSNIF.}
    \label{fig:dynamiclight}
\end{figure*}

To set up a scene with LSNIF, we developed a pipeline that takes a scene description for our reference path tracer, goes through it, performs LSNIF training or conversion for tagged objects, and then writes another scene description for the LSNIF integrator (shown in the supplemental video).
When the converted scene is rendered, we can do some scene editing, such as changing lighting conditions, moving geometries (Fig.~\ref{fig:dynamiclight}), and camera change (Fig.~\ref{fig:dxr}).
Users can also add a pre-trained LSNIF object to a scene as well because LSNIF is optimized for a single geometry and supports its transformation.
Previous methods had restrictions on scene editing, making them a no-go for practical use~\cite{NIF, NBVH}.
To test the robustness of LSNIF and the conversion pipeline, we took scenes designed for other path tracers, such as \textsc{Classroom} and \textsc{Junkshop}, and evaluated the rendered images.
The only change we made for these scenes was small modifications to the materials so that we could load them in our renderer.
\textsc{Classroom} uses geometry instancing which works well with LSNIF as its neural representation is per object, not for the entire scene.
That is, we do not need to train LSNIF for instanced geometries, and the same LSNIF can be used for all instances which reduces the number of LSNIF to be trained from $50$ to $15$.
\textsc{Junkshop} contains various objects that are grouped and converted to LSNIF as shown in Fig.~\ref{fig:junkshopparts}.

When we compute LSNIF, the user can choose the granularity of LSNIF, i.e., how many objects a single LSNIF contains.
The errors are increased as we make it bigger, i.e., creating one LSNIF for more objects.
Fig.~\ref{fig:nifgranularity} is a visual comparison of this parameter where an LSNIF is created for one object or multiple objects.
There are $120$ LSNIF objects in Fig.~\ref{fig:nifgranularity} (a) while there are only eight LSNIF objects in Fig.~\ref{fig:nifgranularity} (c).
The granularity of LSNIF shows a trade-off between memory consumption and quality.
The memory footprint in Fig.~\ref{fig:nifgranularity} (c) is $15\times$ smaller than in Fig.~\ref{fig:nifgranularity} (a) as instancing is turned off for this example.
However, its FLIP error is $1.57\times$ higher.

\begin{table*}[t]
    \scriptsize
    \centering
    \caption{Memory footprint and performance comparison sorted by the number of LSNIF objects. LSNIF count shows the total number of LSNIF objects while values in parentheses exclude instanced geometries. LSNIF, BVH A, and BVH B are the memory footprint of LSNIF, the standard uncompressed BVH, and the estimate of the compressed BVH, respectively. The second row in these columns indicates the ratio to LSNIF in {\color{orange} orange}. They do not include triangles. $T_{LSNIF}^{L}$ and $T_{LSNIF}^{H}$ show rendering times per frame using low and high-quality LSNIF, and their ratios to rendering times with path tracing, $T_{PT}$ (written in {\color{cyan} blue} in the second row). FLIP$^L$ and FLIP$^H$ are the errors for low and high-quality LSNIF compared to path-traced reference images. }
    \label{tab:perf}
    \begin{tabular}{c||cccccccccc}
    Scene & LSNIF count & LSNIF & BVH A & BVH B & $T_{LSNIF}^{L}$ & $T_{LSNIF}^{H}$ & $T_{PT}$ & FLIP$^{L}$ $\downarrow$ & FLIP$^{H}$ $\downarrow$\\
    \hline
    \makecell{\textsc{Hairballs}\\ 5.7M (Tris)} & 2 (1) & \makecell{1.56 MB\\{\color{orange} 1.00}} & \makecell{0.8 GB\\{\color{orange} 525.13}} & \makecell{0.16 GB\\{\color{orange} 106.20}} & \makecell{45.56 ms\\{\color{cyan} 0.54}} & \makecell{65.62 ms\\{\color{cyan} 0.78}} & \makecell{84.20 ms\\{\color{cyan} 1.00}} & 0.115 & 0.066 \\
    \makecell{\textsc{Dragons}\\ 2.6M (Tris)} & 3 (1) & \makecell{1.56 MB\\{\color{orange} 1.00}} & \makecell{0.4 GB\\{\color{orange} 262.56}} & \makecell{0.07 GB\\{\color{orange} 48.44}} & \makecell{39.12 ms\\{\color{cyan} 1.06}} & \makecell{59.61 ms\\{\color{cyan} 1.61}} & \makecell{36.95 ms\\{\color{cyan} 1.00}} & 0.049 & 0.034 \\
    \makecell{\textsc{Statues}\\ 18.2M (Tris)} & 4 (4) & \makecell{6.24 MB\\{\color{orange} 1.00}} & \makecell{2.5 GB\\{\color{orange} 410.26}} & \makecell{0.52 GB\\{\color{orange} 84.77}} & \makecell{37.95 ms\\{\color{cyan} 1.73}} & \makecell{45.15 ms\\{\color{cyan} 2.06}} & \makecell{21.96 ms\\{\color{cyan} 1.00}} & 0.020 & 0.021 \\
    \makecell{\textsc{Junkshop}\\ 20.5M (Tris)} & 9 (9) & \makecell{14.04 MB\\{\color{orange} 1.00}} & \makecell{2.9 GB\\{\color{orange} 211.51}} & \makecell{0.58 GB\\{\color{orange} 42.44}} & \makecell{42.27 ms\\{\color{cyan} 1.07}} & \makecell{49.61 ms\\{\color{cyan} 1.25}} & \makecell{39.54 ms\\{\color{cyan} 1.00}} & 0.035 & 0.037 \\
    \makecell{\textsc{Botanic Cornell Box}\\ 27.0M (Tris)} & 26 (15) & \makecell{23.40 MB\\{\color{orange} 1.00}} & \makecell{3.8 GB\\{\color{orange} 166.29}} & \makecell{0.77 GB\\{\color{orange} 33.54}} & \makecell{150.31 ms\\{\color{cyan} 0.93}} & \makecell{173.97 ms\\{\color{cyan} 1.07}} & \makecell{162.40 ms\\{\color{cyan} 1.00}} & 0.142 & 0.108 \\
    \makecell{\textsc{Botanic Cornell Box}\\ \textsc{Glossy} 27.0M (Tris)} & 26 (15) & \makecell{23.40 MB\\{\color{orange} 1.00}} & \makecell{3.8 GB\\{\color{orange} 166.29}} & \makecell{0.77 GB\\{\color{orange} 33.54}} & \makecell{147.35 ms\\{\color{cyan} 0.93}} & \makecell{176.47 ms\\{\color{cyan} 1.11}} & \makecell{158.76 ms\\{\color{cyan} 1.00}} & 0.198 & 0.170 \\
    \makecell{\textsc{Classroom}\\ 2.6M (Tris)} & 50 (15) & \makecell{23.40 MB\\{\color{orange} 1.00}} & \makecell{0.4 GB\\{\color{orange} 17.50}} & \makecell{0.07 GB\\{\color{orange} 3.23}} & \makecell{53.86 ms\\{\color{cyan} 0.94}} & \makecell{62.01 ms\\{\color{cyan} 1.08}} & \makecell{57.36 ms\\{\color{cyan} 1.00}} & 0.051 & 0.036 \\
    \end{tabular}
\end{table*}

\subsection{Performance Evaluation}
Table~\ref{tab:perf} shows the memory footprint and the performances of LSNIF for various scenes.
We compare the memory footprint of LSNIF with the standard uncompressed BVH (\textit{BVH A}) and the compressed BVH (\textit{BVH B}) with a BVH compression method~\cite{compBVH}.
Note that \textit{BVH A} is a 4-wide BVH used in our performance evaluation, where the size of each internal node is $128$ bytes and the leaf node stores up to two triangles, and the memory footprint for \textit{BVH B} is an estimate using the parameters in the paper. 
LSNIF demonstrates a substantially smaller memory footprint compared to both BVHs, achieving a reduction in size up to $525.1 \times$ and $106.2 \times$ for \textit{BVH A} and \textit{BVH B}, respectively, particularly in complex scenes with a high number of triangles.

In our performance comparison against classical path tracing using BVHs, we tested two configurations of LSNIF: one with a high-quality configuration and the other with a low-quality configuration. Rays are traced up to four bounces in all configurations. 
The high-quality one is our default configuration described in Sec.~\ref{sec:arch}, and the low-quality one is the same as the high-quality one except for the network width and $H$, which are reduced to $64$ and $10$, respectively.
BVH-based path tracing is implemented in a single kernel in our experiments while LSNIF utilizes a wavefront path tracer involving multiple kernel execution.
This leads to higher memory read/write overhead in the LSNIF implementation, which negatively impacts performance compared to single-kernel path tracing.
However, with the high-quality configuration, LSNIF performs better than path tracing in \textsc{Hairballs} though it incurs a rendering overhead of $1.07 \text{--} 2.06\times$ in other scenes.
The low-quality configuration can reduce the performance gap between LSNIF and path tracing.
It achieves better performances than path tracing in \textsc{Botanic Cornell Box}, \textsc{Classroom}, \textsc{Hairballs}, and \textsc{Botanic Cornell Box Glossy}.
We attribute this to LSNIF's efficiency in dealing with instanced geometries and highly-detailed geometries, such as foliage and thin geometry that are computationally expensive for BVH ray tracing.
Although the low-quality LSNIF exhibits increased errors in most scenes, the rendered images are still visually plausible, as shown in the supplemental document.
It also shows equivalent quality with a larger performance overhead in \textsc{Statues} and \textsc{Junkshop}, where all the objects are relatively simple without glossy reflections.
This demonstrates the flexibility of LSNIF, allowing users to balance quality and performance according to their requirements.

\begin{figure*}
    \centering
    \begin{subfigure}[b]{0.162\textwidth}
        \centering
        \includegraphics[width=\textwidth]{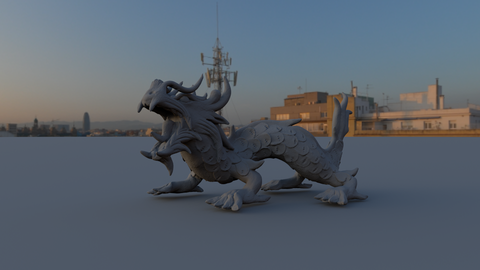}
        \caption{LSNIF}
    \end{subfigure}
    \hfill
    \begin{subfigure}[b]{0.162\textwidth}
        \centering
        \includegraphics[width=\textwidth]{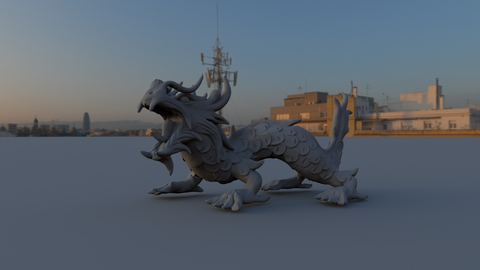}
        \caption{Reference}
    \end{subfigure}
    \hfill
    \begin{subfigure}[b]{0.162\textwidth}
        \centering
        \includegraphics[width=\textwidth]{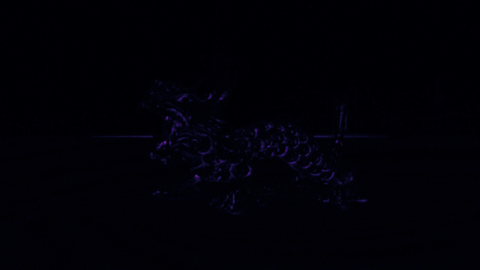}
        \caption{FLIP: $0.008$}
    \end{subfigure}
    \hfill
    \begin{subfigure}[b]{0.162\textwidth}
        \centering
        \includegraphics[width=\textwidth]{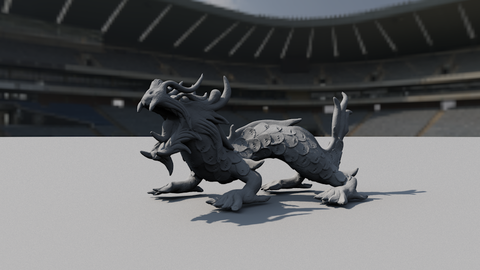}
        \caption{LSNIF}
    \end{subfigure}
    \hfill
    \begin{subfigure}[b]{0.162\textwidth}
        \centering
        \includegraphics[width=\textwidth]{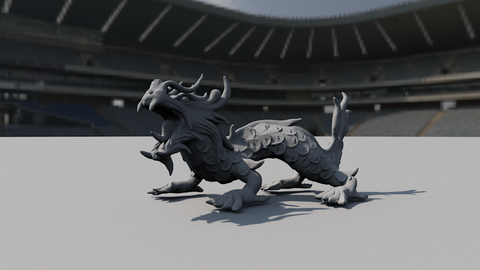}
        \caption{Reference}
    \end{subfigure}
    \hfill
    \begin{subfigure}[b]{0.162\textwidth}
        \centering
        \includegraphics[width=\textwidth]{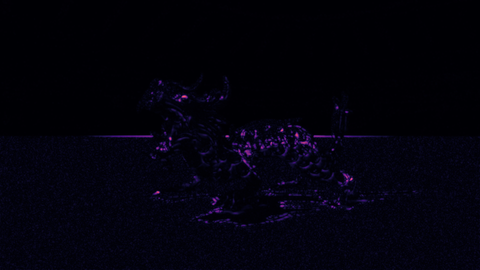}
        \caption{FLIP: $0.018$}
    \end{subfigure}
    \caption{LSNIF produces lower errors than NIF for both cases (0.008 vs. 0.024 and 0.018 vs. 0.031).}
    \label{fig:nifcomp}
\end{figure*}

\begin{figure*}[tb]
    \setlength{\fboxrule}{0.002\linewidth}
    \setlength{\fboxsep}{0\linewidth}
    \centering
    \begin{tabular}{c@{\hspace{0.002\linewidth}}c@{\hspace{0.002\linewidth}}c@{\hspace{0.002\linewidth}}c@{\hspace{0.002\linewidth}}c}
        \raisebox{-0.515\height}[0pt][0pt]{\includegraphics[width=0.421\textwidth]{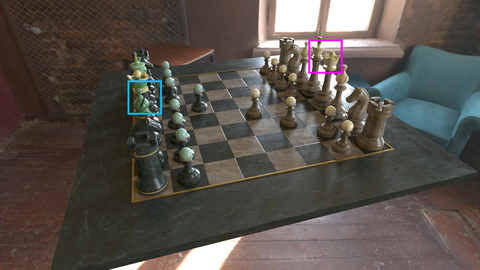}} &
        \raisebox{-0.015\height}[0pt][0pt]{\includegraphics[width=0.208\textwidth]{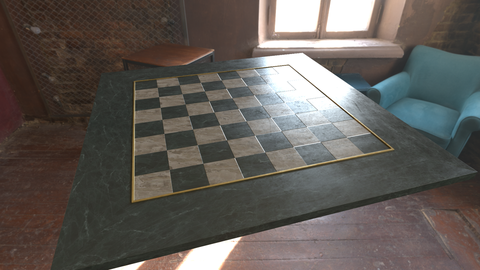}} &
        \fcolorbox{cyan}{cyan}{\includegraphics[width=0.113\textwidth]{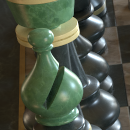}} &
        \fcolorbox{cyan}{cyan}{\includegraphics[width=0.113\textwidth]{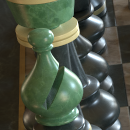}} &
        \fcolorbox{cyan}{cyan}{\includegraphics[width=0.113\textwidth]{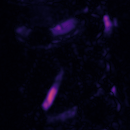}} \vspace{-0.175em} \\

        &
        \raisebox{-0.02\height}[0pt][0pt]{\includegraphics[width=0.208\textwidth]{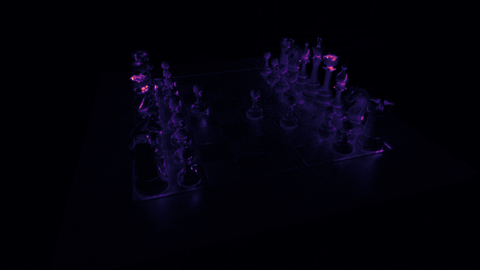}} &
        \fcolorbox{mypink}{mypink}{\includegraphics[width=0.113\textwidth]{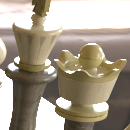}} &
        \fcolorbox{mypink}{mypink}{\includegraphics[width=0.113\textwidth]{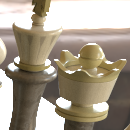}} &
        \fcolorbox{mypink}{mypink}{\includegraphics[width=0.113\textwidth]{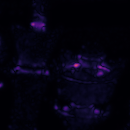}} \vspace{-0.175em} \\

        &
        &
        {\footnotesize LSNIF} &
        {\footnotesize Reference} &
        {\footnotesize FLIP: $0.007$} \vspace{-0.225em}
    \end{tabular}
    \caption{ LSNIF shows a lower error than N-BVH in similar scene settings (0.007 vs. 0.019). \textsc{Chess} contains 12 LSNIF objects consuming smaller memory (19 MB vs. 37 MB). }
    \label{fig:nbvh}
\end{figure*}

\subsection{Comparison with Related Approaches}
\paragraph{NIF~\cite{NIF}}
Since the authors did not release the scenes used in the paper, we created some similar test scenes to compare LSNIF with NIF, as shown in Fig.~\ref{fig:nifcomp}. We selected two tests with low and high-frequency lighting conditions.
The results show that LSNIF produced lower errors for both test cases, outperforming NIF, even when NIF overfits the current scene settings. The FLIP errors with LSNIF are $0.008$ and $0.018$ while the ones with NIF are $0.024$ and $0.031$, respectively.

\paragraph{N-BVH~\cite{NBVH}}
As the renderers are different, we could not reproduce the same images to be compared to N-BVH.
However, we loaded one of the scenes from the N-BVH paper (\textsc{Chess}), adjusted the scene, and evaluated the error of our method with the scene in similar settings.
Here, LSNIF also shows lower errors for this comparison, where the FLIP error with LSNIF is $0.007$, while the error reported in the N-BVH paper is $0.019$.
Out of the $32$ chess pieces, there are some objects for which we can use the instancing technique. The number of LSNIF that we trained for this scene was $12$, thus LSNIF consumes only $18.7$ MB, while N-BVH requires $37$ MB.

\paragraph{Geometric LODs}
These techniques are commonly used to simplify scene geometry and enhance runtime performance. However, simply replacing high-polygon models with lower-polygon models for secondary rays in ray tracing can introduce visual artifacts. Achieving artifact-free rendering requires an advanced traversal algorithm or a mechanism to smoothly transition from fine to coarse LOD, which is not well-supported by current ray tracing APIs. To facilitate a fair comparison between LSNIF and geometric LODs, we substituted the scene's original geometries with coarser LODs, ensuring their BVH sizes matched those of LSNIF. While the coarser LODs work adequately for geometries with simple topologies, they fail to represent the complex topological features of highly detailed geometries accurately, as demonstrated in the supplemental document.

\subsection{Implementation in DirectX Ray Tracing}
\label{sec:dxr}
\begin{figure*}
     \centering
     \begin{subfigure}[b]{0.245\textwidth}
         \centering
         \includegraphics[width=\textwidth]{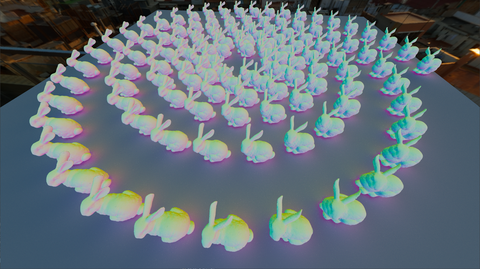}
         \label{fig:pipeline:0}
     \end{subfigure}
     \hfill
     \begin{subfigure}[b]{0.245\textwidth}
         \centering
         \includegraphics[width=\textwidth]{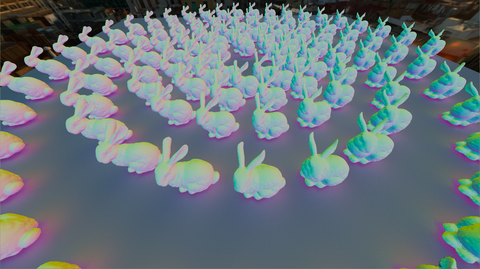}
         \label{fig:pipeline:1}
     \end{subfigure}     
     \hfill
     \begin{subfigure}[b]{0.245\textwidth}
         \centering
         \includegraphics[width=\textwidth]{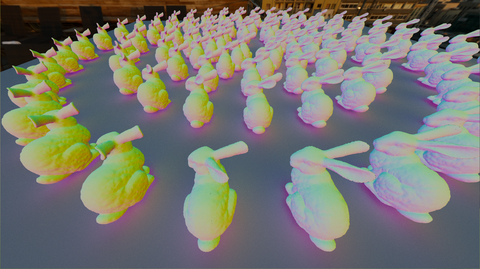}
         \label{fig:pipeline:1}
     \end{subfigure}     
     \hfill
     \begin{subfigure}[b]{0.245\textwidth}
         \centering
         \includegraphics[width=\textwidth]{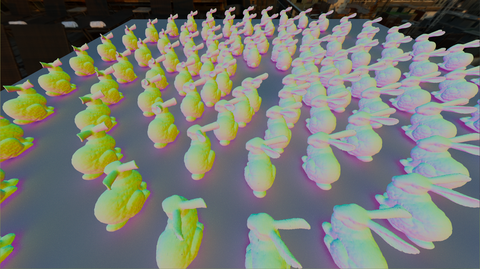}
         \label{fig:pipeline:1}
     \end{subfigure}
     \vspace{-2\baselineskip}
    \caption{LSNIF objects rendered using the DirectX Ray Tracing API and Image-Based Lighting (IBL). The shader invokes inference whenever there is a hit between a ray and the AABB of an LSNIF object. It supports both transformations of the objects as well as camera movements. }
    \label{fig:dxr}
\end{figure*}

The results presented previously were rendered using customized software utilizing the GPU.
However, LSNIF can also be used in a conventional graphics rendering pipeline.
To demonstrate this, we implemented a renderer that features the DirectX ray tracing API, the industry-standard ray tracing API~\cite{dxr}.
An LSNIF object can be implemented as an intersection shader in which the inference is executed to check the intersection of a ray to the object.
Fig.~\ref{fig:dxr} shows the rendered images with the DXR renderer where LSNIF is used for both primary visibility and other visibilities, thus rasterization is not used for this example.
The renderer only has the bunny in LSNIF, and there are no BVH, vertex, and index buffers.
As of today, the performance is not great as we need to execute inference in the intersection shader which causes data and code divergence.
A possible way, also one of the future works, is to improve this using the work graph API which helps to gather coherency and batch the execution~\cite{workgraph}.

\section{Conclusions and Future Work}
In this paper, we introduced LSNIF, which compresses the geometry into neural representation, serving as a substitute for the corresponding part of BVHs in a path tracer.
Trained offline and independently of viewpoints and lighting conditions, LSNIF allows for scene editing, such as changes to lighting, camera positions, and geometry transformations, while maintaining plausible visual quality.
Pre-trained models of LSNIF can be seamlessly integrated and rendered in the scene like standard geometries without additional training.
Although our current implementation does not yet match the speed of hardware-accelerated BVH-based ray tracing, we anticipate narrowing this gap with further optimization and advancements in hardware for matrix operations.
By employing NNs for efficient ray queries at runtime, we have essentially shifted the use of BVH-based ray tracing from runtime to training time.

To minimize rendering error, LSNIF relies on rasterization for primary visibility, which requires retraining vertex and face information of geometries.
However, LSNIF can also be used for primary visibility calculations for applications where primary visibility accuracy is less critical, as shown in the DXR
implementation (Sec.~\ref{sec:dxr}).

There are several limitations to the proposed method.
Firstly, although LSNIF can support some scene modifications, it cannot handle topological changes or deformations of geometry without additional training.
Additionally, LSNIF does not predict texture coordinates, restricting the use of textures in materials, such as normal mapping.
Finding a stable training method for error-sensitive texture coordinates remains challenging and should be a focus for future extensions.
Another area for improvement is the accuracy of predicted geometric properties, which is necessary to completely eliminate mesh representation or rasterization.
The requirement for large input vectors, derived from concatenated feature vectors at multiple intersection points, leads to the use of a large network, which negatively impacts inference speed.
Future research should explore more efficient input vector representations to accelerate inference.
Moreover, the proposed rendering pipeline inherits the limitations of rasterization, such as restrictions on camera types.
Extending the method to support other camera types than perspective cameras, such as spherical or fisheye cameras, would be another future direction.
Currently, all objects represented by LSNIF are assumed to be opaque, and we aim to extend our work to support transmissive materials and sub-surface scattering.
Also, to avoid the self-intersection, we offset the ray origin inferred by LSNIF with the small fixed epsilon value.
Although the artifacts are not shown up in the scene we tested, exploring an efficient way to choose a robust epsilon value adaptively is one of our future works.
Finally, our method does not support levels of detail (LOD) for geometries. Using different voxel resolutions and storing multiple LSNIF models for different LODs is a simple solution but inefficient in terms of memory usage. Further analysis is required to develop a more efficient LOD representation for LSNIF.

\begin{acks}
We are grateful to Piotr Maciejewski and Shikhali Shikhaliev for their help with the implementations.
We thank Nikolai Nikiforov for his help in designing the scenes we rendered in the project.
We also thank Alex Treviño for the \textsc{Junkshop} scene and the Stanford Computer Graphics Laboratory for \textsc{Stanford Bunny}, \textsc{Dragon}, \textsc{Asian Dragon}, and \textsc{Thai Statue} models.
\end{acks}

\bibliographystyle{ACM-Reference-Format}
\bibliography{main}
\end{document}